# Open vs Closed-ended questions in attitudinal surveys – comparing, combining, and interpreting using natural language processing

Vishnu Baburajan [a, b], João de Abreu e Silva [c], Francisco Camara Pereira [d]

[a] PhD Student, MIT-Portugal Program, CERIS, Instituto Superior Técnico, Lisboa, 1049-001, Portugal. Tel: +351 21 841 9865, Email: vishnu.baburajan@tecnico.ulisboa.pt

[b] Guest PhD Student, DTU Management Engineering, Technical University of Denmark, Bygningstorvet, 2800 Kgs., Lyngby, Denmark. Email: vibabu@dtu.dk

[c.] Associate Professor, CERIS, Instituto Superior Técnico, 1049-001, Portugal. Tel: +351 21 841 2302, Email: jabreu@tecnico.ulisboa.pt

[d.] Professor, DTU Management Engineering, Technical University of Denmark, Bygningstorvet, 2800 Kgs., Lyngby, Denmark. Tel: +45 45 25 14 96, Email: camara@dtu.dk

**Corresponding Author**: Vishnu Baburajan [a, b]

## Abstract

To improve the traveling experience, researchers have been analyzing the role of attitudes in travel behavior modeling. Although most researchers use closed-ended surveys, the appropriate method to measure attitudes is debatable. Topic Modeling could significantly reduce the time to extract information from open-ended responses and eliminate subjective bias, thereby alleviating analyst concerns. Our research uses Topic Modeling to extract information from open-ended questions and compare its performance with closed-ended responses. Furthermore, some respondents might prefer answering questions using their preferred questionnaire type. So, we propose a modeling framework that allows respondents to use their preferred questionnaire type to answer the survey and enable analysts to use the modeling frameworks of their choice to predict behavior. We demonstrate this using a dataset collected from the USA that measures the intention to use Autonomous Vehicles for commute trips. Respondents were presented with alternative questionnaire versions (open- and closed-ended). Since our objective was also to compare the performance of alternative questionnaire versions, the survey was designed to eliminate influences resulting from statements, behavioral framework, and the choice experiment. Results indicate the suitability of using Topic Modeling to extract information from open-ended responses; however, the models estimated using the closed-ended questions perform better compared to them. Besides, the proposed model performs better compared to the models used currently. Furthermore, our proposed framework will allow respondents to choose the questionnaire type to answer, which could be particularly beneficial to them when using voice-based surveys.

## 1. Introduction and background

For over a century, researchers and policymakers have been interested in understanding the psychological constructs characterizing an individual's attitudes. In their pursuit of a better understanding of human behavior, researchers have measured attitudes in various areas of research, such as psychology (Plant, 1922), politics (Buccoliero et al., 2020; Geer, 1991) and medicine (Amin, 2020; Sun et al., 2020). In travel behavior research, attitudinal constructs have been used extensively in the analysis of the intention to use new services (Payre et al., 2014), preferences (Kaplan et al., 2014), satisfaction (Ge et al., 2017) - to name a few.

In this context, one can use open-ended (e.g., What is your opinion of the inflight experience?) or closed-ended questions (e.g., On a scale of 1-5, rate your inflight experience) to measure attitudes. And researchers have long been divided on which of these is best suited to measuring attitudes. Driven primarily by convenience, having a simpler process for respondents, analysts and policymakers, and a swifter operation (Bowling, 2005; Converse, 1984), the majority of the researchers have used closed-ended questions to measure attitudes. The additional time needed to complete the respective questionnaire and make inferences from open-ended surveys (Krosnick, 2018) has also contributed to this.

A set of statements describing attitudes and a scale is presented to the respondents in the closed-ended survey. And they respond by choosing points on the scale that best represents their attitude. These scales can be bipolar (e.g., a 5-point scale ranging between "Strongly disagree" and "Strongly agree") or unipolar (e.g., a 5-point scale ranging between "Not at all satisfied" and "Extremely satisfied"); the former is predicated on the notion that attitudes are bipolar constructs. The latter assumes a unipolar construct measuring only the intensity of the attitude (Krosnick and Fabrigar, 1997). Despite the previously mentioned advantages of the closed-ended approach, such an approach ignores the cognitive burden on the respondents described by Tourangeau and Rasinski (1988), as respondents go through a series of steps – a. *interpret the question*; b. *recognize the attitude being measured*; c. *recollect their beliefs and feelings;* and d. *relate the attitudes to the point on the scale that best describes their attitudes*. Furthermore, the researcher must carefully decide on the type of scale (unipolar/bipolar), length, labelling, the inclusion of mid-points, etc., before using the scale, some of which are still a subject of intense debate among researchers (Krosnick and Fabrigar, 1997). Excluding the middle (neutral) option might help reduce the number of respondents choosing the middle option (Bowling, 2005) and avoid social desirability bias, i.e., respondents choosing answers that make them look good in the eyes of the analysts (Garland, 1991). But it may also force individuals with a neutral opinion to agree/disagree, making it more challenging to measure neutral opinions (Matell and Jacoby, 1972). The closed-ended approach also suffers from acquiescence – i.e., respondents endorsing statements irrespective of their content (Bowling, 2005). The advent of online surveys has added additional challenges for researchers, as respondents might be more sensitive to scales (Wells et al., 2014) and their visual presentation [vertical/horizontal] (Galesic et al., 2008; Weijters et al., 2020). Researchers have used scales of varying lengths (from 2 points to 101 points) to measure attitudes (Bansal et al., 2016; Buckley et al., 2018; Nishihori et al., 2017; Nordhoff et al., 2018; Preston and Colman, 2000; Qu et al., 2016). And to extract information from closed-ended responses, they have used Exploratory Factor Analysis using Principal Components (Baburajan and de Abreu e Silva, 2019), Exploratory Factor Analysis (Zhang, 2019) and Confirmatory Factor Analysis (Elizabeth et al., 1995), depending on the objectives of the research.

In contrast, the open-ended approach allows respondents to express their attitudes freely in their own words and encourages them to be expressive without being biased or bound to closed alternatives (Converse, 1984), thereby reducing the respondents' cognitive burden. These characteristics also make open-ended questions quite effective when measuring knowledge while producing more reliable and valid results (Krosnick and Presser, 2010). Their use is critical in complicated and relatively new issues that might attract very little attention or problems people may not have thought about extensively (Rugg and Cantril, 1942). However, the use of open-ended questions also poses severe challenges for respondents, enumerators and analysts (Lazarsfeld, 1944). The critics of the open-ended approach frequently argue that they measure merely the ability to articulate a response and not the attitude (Stanga and Sheffield, 1987). Moreover, while designing an open-ended question, researchers should carefully design the space for the response and provide self-explanatory instructions, as these can affect the length and quality of the responses (Smyth et al., 2009). Additionally, open-ended surveys are also associated with higher costs, the vagueness of responses (Converse, 1984; Krosnick, 2018), higher item non-response (Holland and Christian, 2009) and survey break-offs (Peytchev, 2009). Mainly from the angle of qualitative analysis and theme coding, researchers have used open-ended questions to measure perceptions of real-time transportation information displays (Ge et al., 2017) and attitudes towards cycling (Manaugh et al., 2017; Underwood et al., 2014). In these studies, analysts first identified the themes so as to devise the coding template, which coders then used to extract information from open-ended responses. Coders refined the template until they found an acceptable level of reliability in the coded responses between the various coders. Deutsch and Goulias (2012) used open-ended questions to understand differences between places and semantic frequency analysis to extract the themes. More sophisticated approaches such as word-based (Jehn, 1995; Niedomysl and Malmberg, 2009) or content-based (Hsieh and Shannon, 2005; Mamali et al., 2020) analysis, which relies on the frequency of occurrence of words in the text, have also been used. Other methods used to extract information include Automated Text Analysis (Mossholder et al., 1995; ten Kleij and Musters, 2003) and Topic Models (Baburajan et al., 2018; Mitsui et al., 2020; Roberts et al., 2014; Tvinnereim and Fløttum, 2015).

Now that we have presented the two approaches, their merits, and their demerits, we would stress that some of the concerns related to using the two approaches discussed above remain unresolved. For instance, although there are some valid arguments in favor of open-ended responses, there are challenges related to extracting information and drawing inferences from them, making the use of the closed-ended approach more attractive to an analyst/researcher. Furthermore, it can be argued that the respective studies focused excessively on the convenience of the analyst/researcher and not specifically on that of the respondent, it is highly plausible to assume that not every respondent might favor one of these approaches over the other. Considering this, our research pursues the following objectives: a. *to develop a framework that equips analysts to allow respondents to use the questionnaire type of their choice by using established models to predict the attitudes of interest to them;* b. *to compare and benchmark the performance of the proposed model with established models currently used*.

Our previous study (Baburajan et al., 2020) analyzed the suitability of using Topic Modeling approaches to extract information from open-ended responses, which we also reanalysed in this paper. However, *the primary contribution of this paper is the proposed framework that enables analysts to use different questionnaire types to measure attitudes yet use the existing models to draw inferences and make predictions*. This framework will also be a step towards making data collection more flexible, as respondents can then choose the questionnaire type convenient to them. Moreover, one could reap the most benefits by incorporating voice-based surveys and speech recognition tools. For instance, some individuals may find it convenient to answer surveys quickly by talking, rather than writing or choosing some points on a scale. In such circumstances, analysts could use our proposed framework to model attitudes. To accomplish the two objectives of this paper, we develop an experimental design that presents alternative questionnaire versions to the respondent (discussed in Section 2). Having collected the data, we evaluated the representativeness of the data and performed the exploratory analysis, which is presented in Section 3. In line with the first objective, we then expand on the framework that allows respondents to answer the survey more flexibly using Probabilistic Graphical Models (PGMs), as described in greater depth in Section 4. In Section 5, we discuss the extraction of information from both the open- and closed-ended questions. The estimation results and a discussion of the results follow in Section 6. A mapping of responses based on the attitudes using our proposed framework between the different questionnaire versions features is presented in the penultimate section. The final section summarizes our research findings and limitations thereof.

## 2. Experimental design and data collection

We collect data by means of an online survey that investigates the role of attitudes in the intention to use Autonomous Vehicles (AVs) for commute trips, using a Stated Preference (SP) experiment. We narrow down the influence of this intervention to the questionnaire type by controlling for the effect of statements, behavioral framework, and choice experiments. The behavioral framework and statements are adopted from Zhang et al. (2019), which is an extension of the Technology Acceptance Model (Davis, Jr, 1985), together with an SP experiment developed by Haboucha et al. (2017).

Findings from our previous research emphasize that placing open-ended questions before the closed-ended questions influences the responses to the closed-ended questions (Baburajan et al., 2020). Considering this, we used three questionnaire versions (Ver_LK, Ver_LKOE and Ver_OE). Ver_LK used only five-point Likert scale questions, whereas Ver_OE used only open-ended questions to measure attitudes. In Ver_LKOE, two open-ended questions were used before the Likert scale questions, and the responses to these open-ended questions were not used to measure attitudes. Our sole intention in using these questions was to prime the Likert scale responses. We used the following two questions: 1. *What are your general opinions about Autonomous Vehicles? (open-ended question on Attitudes in Ver_OE);* and 2. *Do you believe that Autonomous Vehicles are useful? Explain why (open-ended question on Perceived Usefulness in Ver_OE).* In all three questionnaires, attitudes related to the "Perceived Ease of Use (PEoU)", "Perceived Usefulness (PU)", "Perceived Safety Risk (PSR)", "Perceived Privacy Risk (PPR)", "Trust (Tr)" and "Attitudes towards AVs (Att)" were measured.

Our estimation results did not indicate a significant improvement in performance using this framework over the model that used only PU and PEoU; for the sake of model parsimony, we limit our discussion of attitudes in this paper to PU and PEoU. The statements used along with the open-ended questions are presented in Table 1. In addition to these, the questionnaire collected information on characteristics such as socio-demographics, travel, familiarity with AVs [based on (US Department of Transportation, 2018)] and finally, mode choice for commute trip (using an SP experiment). We collected data by presenting alternative questionnaire versions (incorporating closed- and open-ended questions) to the respondents using the Randomization feature of Qualtrics (Qualtrics, 2021).

**Table 1**

Statements for Likert scale questions and open-ended questions

| Psychological Construct | Statements | Open-ended Questions |
|---|---|---|
| Perceived Ease of Use | Cronbach's alpha | |
| (PEoU) | [Ver_LK- 0.905, Ver_LKOE- 0.908] | Do you think that it will be easy to use Autonomous Vehicles? Explain why. |
| | Learning to use autonomous vehicles will be easy for me | |
| | I will find it easy to get autonomous vehicles to do what I want it to do | |
| | It will be easy for me to become skillful at using autonomous vehicles | |
| | I will find autonomous vehicles easy to use | |
| Perceived Usefulness | Cronbach's alpha | |
| (PU) | [Ver_LK- 0.830, Ver_LKOE- 0.841] | Do you believe that Autonomous Vehicles are useful? Explain why. |
| | Using autonomous vehicles will be useful in meeting my driving needs | |
| | Autonomous vehicles will let me do other tasks, such as eating, watch a movie, be on a cell phone on my trip | |
| | Using autonomous vehicles will decrease my accident risk | |
| | Using autonomous vehicles will relieve my stress of driving | |
| | I find autonomous vehicles to be useful when I'm impaired (e.g., drowsy, drunk, drugs) | |

We launched the survey in the United States of America between January and March 2020 and collected responses from the survey panels provided by Cint (Cint, 2021). Cint contacted respondents on our behalf, so we did not have direct access to the respondents during and after the completion of the study. However, in the introduction to the survey, we presented the researchers' names and affiliations, allowing respondents to seek clarifications or request findings from our research. In addition to this, the respondents were informed that the researchers would ensure data anonymity and use the collected data exclusively for research without sharing it with third parties.

Respondents answering too casually or inconsistently were removed from the dataset and replaced by Cint. To deal with respondents answering the questionnaire too quickly (we refer to them as "speeders"), we computed the minimum time for a respondent to answer the different questions in our survey, using the information from Qualtrics (2020). Individuals who responded faster than this minimum stipulated time were classified as "speeders". We omitted respondents who answered in less than seven words for the open-ended responses, as we wanted respondents to use a full sentence to answer the question. Opting for a higher value for the optimal number of words in a sentence would significantly burden the team and our panel provider Cint, as they replaced respondents that did not adhere to the norms. Hence, considering the optimal number of words in a sentence[1], we opted for seven words. These responses were removed only after data collection, as enforcing conditions while answering the questionnaire could alter the actual response. The final dataset representing the US population based on gender, ethnicity and regional diversity comprizes 3002 complete responses (Ver_Lk - 1012, Ver_LkOE - 1021, Ver_OE - 969).

---

[1] https://medium.com/@scottydocs/what-is-the-perfect-sentence-length-4690ce8d5048

## 3. Data description

### 3. 1. Socio-demographic characteristics

We evaluated the differences in the socio-demographic characteristics of the respondents answering the alternative questionnaire versions (presented in Table 2). Nearly 53% of respondents were female, and the average age was 38.69 (std. dev: 13.91). Ver_LKOE and Ver_OE of the questionnaire had a slightly higher portion of respondents earning between $50,000 and $74,999 and possessing a bachelor's/graduate degree. Almost 70% of the respondents were European American, and 14-18% were African American. Moreover, approximately 55% of the participants were employed full-time, 30% part-time, and Ver_OE had slightly more students. As expected, the results of our Mann-Whitney U (Mann and Whitney, 1947) test indicated that the differences were not statistically significant for gender and ethnicity, but they were statistically significant for other variables (see Table 3). Accordingly, in the estimation of the models, we also include these socio-demographic variables as explanatory variables.

**Table 2**

Socio-demographic characteristics of respondents in the survey.

|  |  | Frequency | | |
| --- | --- | --- | --- | --- |
| Variable | Levels | Ver_LK | Ver_LKOE | Ver_OE |
| Gender (%) | Female | 52.87 | 53.18 | 53.66 |
|  | Male | 46.74 | 46.43 | 46.13 |
|  | Prefer not to answer | 0.4 | 0.39 | 0.21 |
| Household income (%) | Less than $10,000 | 9.39 | 7.74 | 5.68 |
|  | $10,000 - $14,999 | 5.14 | 3.82 | 2.89 |
|  | $15,000 - $24,999 | 10.87 | 8.81 | 7.84 |
|  | $25,000 - $34,999 | 12.85 | 12.54 | 12.69 |
|  | $35,000 - $49,999 | 15.51 | 15.38 | 15.48 |
|  | $50,000 - $74,999 | 17.19 | 19.78 | 20.64 |
|  | $75,000 - $99,999 | 12.06 | 12.14 | 13.93 |
|  | $100,000 - $124,999 | 5.24 | 6.56 | 7.33 |
|  | $125,000 - $149,999 | 2.47 | 3.53 | 3.1 |
|  | $150,000 - $199,999 | 2.47 | 2.84 | 2.27 |
|  | More than $200,000 | 1.88 | 1.96 | 2.68 |
|  | Prefer not to answer | 4.94 | 4.9 | 5.47 |
| Educational qualification (%) | Less than high school graduate | 6.23 | 5.48 | 6.91 |
|  | High school graduate or GED | 28.85 | 23.41 | 20.33 |
|  | Some college or associate degree | 38.24 | 38.69 | 36.95 |
|  | Bachelor's degree | 17.29 | 21.94 | 23.74 |
|  | Graduate degree or professional degree (Master or PhD) | 8.2 | 9.79 | 11.76 |
|  | I prefer not to answer | 1.19 | 0.69 | 0.31 |
| Ethnicity (%) | European American | 69.37 | 70.71 | 69.45 |
|  | African American | 17.39 | 14.99 | 13.83 |
|  | American Indian or Alaska Native | 1.19 | 1.47 | 1.86 |
|  | Asian | 4.05 | 3.33 | 5.06 |
|  | Native Hawaiian or other Pacific Islander | 0.59 | 0.49 | 0.72 |
|  | Some other race | 6.62 | 8.23 | 7.53 |
|  | I prefer not to answer | 0.79 | 0.78 | 1.55 |
| Employment status (%) | Full-time | 55.93 | 52.01 | 53.46 |
|  | Part-time | 30.24 | 31.73 | 27.55 |
|  | Student | 13.83 | 16.26 | 18.99 |
|  |  | Average (Std. Dev.) | | |
| Age |  | 40.90 (13.31) | 38.61 (14.23) | 36.55 (14.19) |
| Household size |  | 3.22 (1.83) | 3.12 (1.75) | 3.07 (1.71) |
| Number of adults in the HH |  | 2.23 (0.99) | 2.24 (1.00) | 2.27 (1.00) |
| Number of children aged between 8 and 17 |  | 0.62 (0.97) | 0.57 (0.97) | 0.50 (0.92) |
| Number of children aged under 8 |  | 0.38 (0.78) | 0.31 (0.71) | 0.29 (0.64) |

**Table 3**

Results of Mann-Whitney Analysis (p-values).

| Variable | Ver_LK vs Ver_LKOE | Ver_LK vs Ver_OE | Ver_LKOE vs Ver_OE |
|---|---|---|---|
| Gender (%) | 0.64 | 0.64 | 0.55 |
| Household income (%) | 0.00 | 0.00 | 0.00 |
| Educational qualification (%) | 0.00 | 0.00 | 0.00 |
| Ethnicity (%) | 0.40 | 0.40 | 0.53 |
| Employment status (%) | 0.00 | 0.00 | 0.00 |

We then investigated the frequency distribution of the mode choice for commute trips – the response to the Stated Preference questions (presented in Fig. 1). We observed a difference depending on the questionnaire type used. Of those responding to Ver_LK of the questionnaire, 49.89% of participants preferred "Regular Car", 29.58% "Private AV", and the rest (20.53%) "Shared AV". The distribution of the shares was different for Ver_LKOE but was similar to Ver_OE. Approximately 40% (Ver_LKOE: 40.84% and Ver_OE: 39.15%) of the respondents chose "Regular Car", ≈34% (Ver_LKOE: 34.51% and Ver_OE: 33.55%) chose "Private AV", while the remaining (≈25%) (Ver_LKOE: 24.65% and Ver_OE: 27.3%) chose "Shared AV" for their commute trips. Comparing the distributions, we observe a noticeable difference, as about 50% of respondents chose "Regular Car" in Ver_LK, which, we believe, will have implications for the estimated coefficients and the models' performance. Confirming our belief, the results of analysis for Ver_LK indicates that models perform poorly for the test set, as it overfitted for the dataset (described later in Section 6).

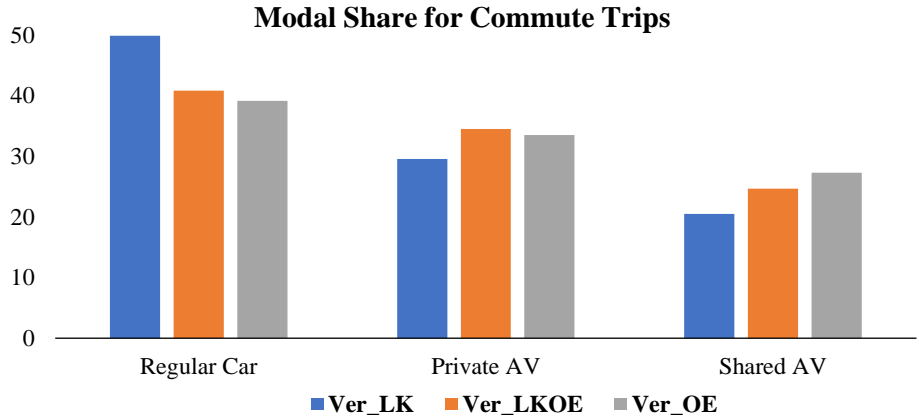

**Fig. 1** Frequency distribution of commute mode choice

We analyzed if these differences in the distribution are statistically significant. We did this by estimating a Multinomial Logit (MNL) model for the mode choice by using "questionnaire type" – an indicator variable for the questionnaire type as the independent variable. For a detailed description of the underlying principles on MNL, readers may consult Ben-Akiva and Lerman (1985). We estimated the model by using "Regular car" as the base alternative, and the questionnaire type was found to be statistically significant at 99% in the utility equations for both "Private AV" (Ver_LKOE - [coef - 0.354, t-stat - 8.39]; Ver_OE - [coef- 0.368, t-stat- 8.54]) and "Shared AV" (Ver_LKOE- [coef- 0.383, t-stat- 8.13]; Ver_OE- [coef- 0.527, t-stat- 11.17]).

## 4. Methodology

### 4. 1. Probabilistic Graphical Models

We applied a Machine Learning modelling framework called Probabilistic Graphical Models (PGMs) for this project. PGMs are comparable to Bayesian Structural Equation Models (BSEMs, (Muthén and Asparouhov, 2012)). Both have their origins in Path Analysis and provide a pictorial representation to depict the relationships between the different variables. Where BSEMs use a single model for the entire system, PGMs leverage conditional independence to fit a sequence of models that model the complete system (Koller and Friedman, 2009; Smith et al., 2009). The use of BSEMs requires a high degree of statistical expertise and the multiple steps involved in the BSEM workflow increases the estimation time significantly (Conrady and Jouffe, 2015). In the case of PGMs, researchers focus on the stochastic processes that generate the data and then use probability theories to account for the uncertainty. Furthermore, the implementation and estimation of PGMs are greatly simplified by the availability of popular tools like Pyro (Bingham et al., 2019) or STAN (Stan Development Team, 2021). In developing a PGM, researchers utilize the following three components (Peled et al., 2019):

1. **Generative Process** – here, we envision the process that generated the given data, which would clarify the assumptions regarding the structure of the uncertainty for the problem.
2. **Probabilistic Graphical Model (graphical representation)** – these are graphs that present the probabilistic models intuitively and compactly. These graphs represent random variables using nodes and the probabilistic relationships using the edges.

3. **Joint Probability Distribution** – one can obtain the marginal distribution of any subset of these variables using the joint probability distribution computed using the sum rule of probability.

In PGMs, researchers can specify probability distributions of their choice for the random variables along with the prior distribution (Airoldi, 2007; Farasat et al., 2015). Taking into consideration priors and likelihood formulations, PGMs fit the data to obtain the distributions of the constants and latent variables. Furthermore, researchers cannot obtain closed-form solutions in most cases, and therefore use approximations obtained using Monte Carlo Markov Chains (MCMC), expectation-maximization (EM) or variational methods (Airoldi, 2007). These are already included in tools like Pyro (Bingham et al., 2019) and STAN (Stan Development Team, 2021).

Let us now look at an example. Below, we show the linear regression model's generative process and joint probability distribution. Fig. 2 shows the PGM graphical representation. The shaded nodes represent the observed variables, and the unshaded nodes represent latent variables. Scalars are denoted using letters in regular fonts (σ, λ), vectors using lower case bold letters (**x**, ***β***) and matrices using upper case bold letters (**X**). In Table 4, we present a comparison between SEMs and PGMs. As shown in Table 4 and Fig. 2, in PGMs, the latent variables and the estimated coefficients are both depicted inside an unshaded node. For instance, in the table, we represent the latent variable "Z" and the coefficient "β" inside unshaded circles. After the coefficients are estimated, they are then represented using shaded nodes. Referring to the example for the linear regression model in Fig. 2., observations $Y_n$ are generated for each value of $X_n$ relying on the coefficients β; hence, the arrows are directed towards $Y_n$, both from $X_n$ and β (referring to eq. 2).

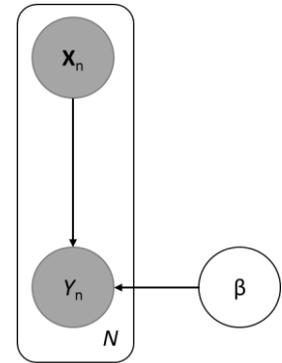

**Fig. 2.** PGM for Linear Regression

**Generative Process**

Given **X**, σ, λ:
1. Draw ***β*** ~ N (**0**, λ **I**) (eq. 1)
2. For each feature vector $\mathbf{x}_n$, $n = 1\ldots N$:
   a. Draw variable $y_n$ ~ N ($\boldsymbol{\beta}^T \mathbf{x}_n$, $\sigma^2$) (eq. 2)

**Joint Probability Distribution**

$p(\boldsymbol{\beta}, \mathbf{y} \mid \mathbf{X}, \sigma, \lambda) = \underbrace{N(\boldsymbol{\beta} \mid 0, \lambda \mathbf{I})}_{prior} \underbrace{\prod_{n=1}^{N} N(y_n \mid \boldsymbol{\beta}^T \mathbf{x}_n, \boldsymbol{\sigma}^2)}_{likelihood}$ (eq. 3)

**Table 4**

Comparison of Structural Equation Models with Probabilistic Graphical Models (Ho et al., 2009; Peled et al., 2019)

| Elements | Structural Equation Models | | Probabilistic Graphical Models | |
|---|---|---|---|---|
| Observed variables | Variables inside Squares | X | Variables inside shaded circles | X |
| Latent Variables | Variables inside unshaded ellipses | (Z) | Variables inside unshaded circles | (Z) |
| Intercepts | Triangles | △1 | Variables inside unshaded circles | (α) |
| Coefficients | | | Coefficients inside unshaded circles | (β) |
| Estimated coefficients | | | Coefficients inside shaded circles | (β shaded) |
| Direct effect | Directional arrow | → | Directional arrow | → |
| Non-directional association between variables | Curved bi-directional arrow | ⌒ | | |

*4. 2. Extracting topics from text using Latent Dirichlet Allocation (LDA)*

LDA is a popular Topic Modeling approach that identifies latent constructs in text data and can capture the information from open-ended responses. It is built on the idea that words specific to a given topic are more likely to occur in documents that discuss the given topic(s). LDA functions like a multinomial principal component analysis (PCA) by converting a text document into a linear combination of topics (word frequencies represent both the text document and topics).

We use a vector of word frequencies to represent a document, often termed the Bag of Words (BoW) representation. For a given number of topics and BoW, LDA then extracts K topics (also a BoW) by minimizing the reconstruction error of the original documents. Thus, projecting a document in the topic space means documents are represented again as a combination of K topics[2]. α and β are the Dirichlet priors for word-topic distributions and document-topic distributions. We will now discuss the PGM in Fig. 3. In the figure, we have a dataset with "D" documents, each containing $N$ words, and we refer to each word using $w_{d,n}$ ($n = 1, ..., N; d = 1, ..., D$). Each word is then assigned to a topic $k$, and each topic $k$ ($k = 1, ..., K$) consists of a vector $β_k$ composed of the words and associated frequencies for this topic "k". A detailed discussion of LDA can be found in Blei et al. (2003), and plenty of accessible tutorials with code exist on the internet today[3,4]. From the LDA, we obtain a set of topics that act as "common building blocks" from the given corpus[5] of documents. And we can obtain knowledge on how much each document belongs to a specific topic to estimate the models for mode choice.

**Generative Process**

For each document **w** in a corpus D:
1. Choose $N \sim \text{Poisson}(ξ)$ (eq. 4)
2. Draw topic $φ_k$ such that $φ_k \sim \text{Dir}(β)$ for $k = 1, ..., K$ (eq. 5)
3. For each document $d$ ($d = 1, ..., D$):
    a. Choose $θ \sim \text{Dir}(α)$ (eq. 6)

---
[2] just like a vector is re-represented as a combination of eigenvectors in PCA
[3] https://towardsdatascience.com/end-to-end-topic-modeling-in-python-latent-dirichlet-allocation-lda-35ce4ed6b3e0
[4] https://www.machinelearningplus.com/nlp/topic-modeling-gensim-python/
[5] a collection or body of knowledge or evidence

b. For each of the *n*-words within the document d, $w_{d,n}$:
     i. Choose a topic assignment $z_{d,n}$ ($z_{d,n} \in \{1, ..., K\}$) such that $z_{d,n} \sim$ Multinomial($\theta_d$) (eq. 7)
     ii. Choose a word $w_{d,n}$ such that $w_{d,n} \sim$ Multinomial($\varphi z_{d,n}$) (eq. 8)

**Joint Probability Distribution**

$$p(\varphi, \theta, z, w) = \left(\prod_{d=1}^{D} p(\theta_d|\alpha) \prod_{n=1}^{N} \left(p(w_{d,n}|\varphi_{z_{d,n}})p(z_{d,n}|\theta_d)\right)\right) \prod_{k=1}^{K} (\varphi_k|\beta) \quad \text{(eq. 9)}$$

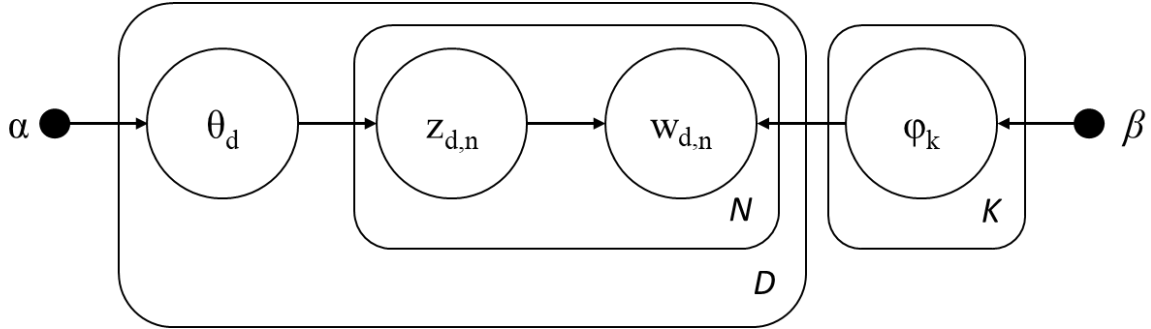

**Fig. 3.** PGM for LDA

α and β are the given parameters that define the priors for $\theta_d$ and $\varphi_k$ respectively. The topic assignment $z_{d,n}$ is generated based on $\theta_d$, and the word $w_{d,n}$ is determined based on its assigned topic ($z_{d,n}$) and the word distribution associated with that topic ($\varphi_{z_{d,n}}$).

*4. 3. PGM models for Mode Choice for Commute Trips*

First, we develop a PGM reflecting the "classic" approach. Each instrument (e.g., Likert scale; open-ended questionnaire) has its independent PGM model developed and estimated separately, which we call *Individual* models. We then propose a combination of the techniques in a single PGM. The intuition is that we ultimately deal with the same behavioral and attitudinal phenomenon regardless of the instrument (closed- or open-ended questionnaires). Therefore, the actual model should be that which can be estimated from the three datasets (Ver_LK, Ver_LKOE, and Ver_OE) simultaneously.

*4. 3. 1 Individual models*

We estimate individual models for each version of the questionnaire. The PGM presented in Fig. 4 is the same for the different types of questionnaires. In Fig. 4:-

1. Att $_{in}$ – attitudes measured using closed-/open-ended questions
2. Att – latent attitudes
3. $X_s$ – socio-demographics and travel characteristics of the individual
4. Y – the observed mode choice for commute trips

The larger plate marked "N" indicates N data observations of the model, and the smaller plate marked "C" indicates C alternatives (Regular car, Private AV, and Shared AV).

We assume that the latent attitudes are identically and independently distributed ("i.i.d") [K-dimensional] with a mean estimated using equation (12) and unit standard deviation. We also assume a normal distribution (mean: 0; std. dev: 1) for the coefficients α$_{in}$ and γ$_{in}$. Considering the nominal nature of the observed choice variable, "mode choice for commute trips (Y)" is presumed to have a multinomial logit formulation with "Regular Car" as the base alternative. We use alternative specific constants (α$_C$) and coefficients (β$_C$) for "Private AV" and "Shared AV". The utility is computed using the socio-demographics, travel, attitudes (Att), and variables of the choice experiment (note that eq. 15 corresponds to a classical MNL).

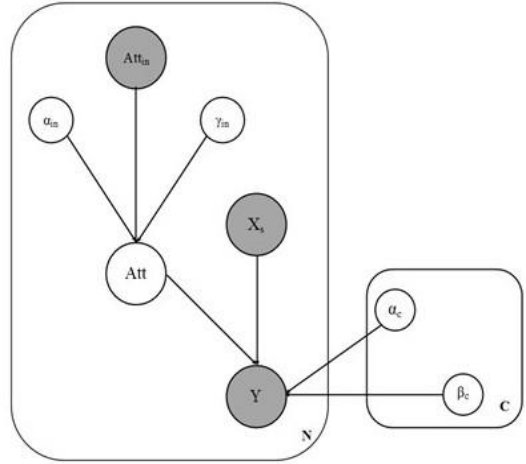

**Fig. 4.** PGM for the individual level model

**Generative Process**

1. For equation $k \in \{1, \ldots, K\}$:
   a. $Draw\ coef\ \alpha_{in\ (k)} \sim Normal\ (\alpha_{in\ (k)} \mid 0, 1)$ (eq. 10)
   b. $Draw\ coef\ \gamma_{in\ (k)} \sim Normal\ (\gamma_{in\ (k)} \mid 0, 1)$ (eq. 11)
   c. $Draw\ coef\ Att_{(k)} \sim Normal\ \left(Att_{(k)} \mid (\alpha_{in\ (k)} + \gamma_{in\ (k)} * Att_{in\ (n)})\right), 1)$ (eq. 12)

2. For each class $c \in \{1, \ldots, C\}$:
   a. $Draw\ coef\ \alpha_c \sim Normal\ (\alpha_c \mid 0, 1)$ (eq. 13)
   b. $Draw\ coef\ \boldsymbol{\beta_c} \sim Normal\ (\boldsymbol{\beta_c} \mid 0, 1)$ (eq. 14)
   c. $Draw\ choice\ \boldsymbol{Y_c} \sim Multinomial\ (\boldsymbol{Y_c} \mid Softmax\ ([\boldsymbol{X_s}, \boldsymbol{A_t}], \alpha_1 \ldots \alpha_C,\ \boldsymbol{\beta_1} \ldots \boldsymbol{\beta_C}))$ (eq. 15)

**Joint Probability Distribution**

$$p(\alpha_{in}, \gamma_{in}, \alpha_C, \beta_C, Att_{in}, \boldsymbol{Att}, \boldsymbol{X_S}, Y)$$
$$= p(\alpha_{in})p(\gamma_{in})(\prod_{c=1}^{C} p(\alpha_c))(\prod_{c=1}^{C} p(\beta_c)) \begin{pmatrix} p(Att_{in}) \\ \prod_{n=1}^{N}\quad p(\boldsymbol{Att} \mid Att_{in}, \alpha_{in}, \gamma_{in}) \\ p(\boldsymbol{X_S})\ p(Y \mid \boldsymbol{Att},\ \boldsymbol{X_S}, \alpha_C, \beta_C) \end{pmatrix} \quad \text{(eq. 16)}$$

*4. 3. 2 Proposed model*

We extend the individual models by combining the attitudes measured using different approaches (closed- and open-ended questions), based on the idea that the same attitude is measured using other methods. In the *Proposed Model* (see Fig. 5):

1. Att $_{lk}$, Att $_{lkoe}$ and Att $_{oe}$ – measured attitudes for Ver_LK, Ver_LKOE and Ver_OE
2. Att – latent attitudes
3. X$_s$ – socio-demographics and travel characteristics of the individual
4. Y – the observed mode choice for commute trips
5. α$_{lk}$, α$_{lkoe}$ and α$_{oe}$ – constants for Ver_LK, Ver_LKOE, Ver_OE
6. γ$_{lk}$, γ$_{lkoe}$ and γ$_{oe}$ – coefficients for Ver_LK, Ver_LKOE, Ver_OE.

Our proposed model formulation assumes a single latent attitude variable (Att) regardless of the instrument used to collect data. The way we have structured these models allows for only one relevant coefficient for the dataset to be estimated at a time (see eq. 23).

This formulation also eliminates the potential for identification issues, as knowing the coefficients for one instrument (e.g., Ver_LK) does not create problems calculating the coefficients for other instruments (e.g., Ver_LKOE and Ver_OE) because these will be "deactivated". The issue related to identification can also be tested experimentally by evaluating the convergence of the coefficients, particularly when using multiple Monte Carlo Markov chains for the estimation. In this regard, researchers often assess the "R-Hat" values for each estimated coefficient, which should ideally be closer to 1 when chains are mixed well (Johnson et al., 2022).

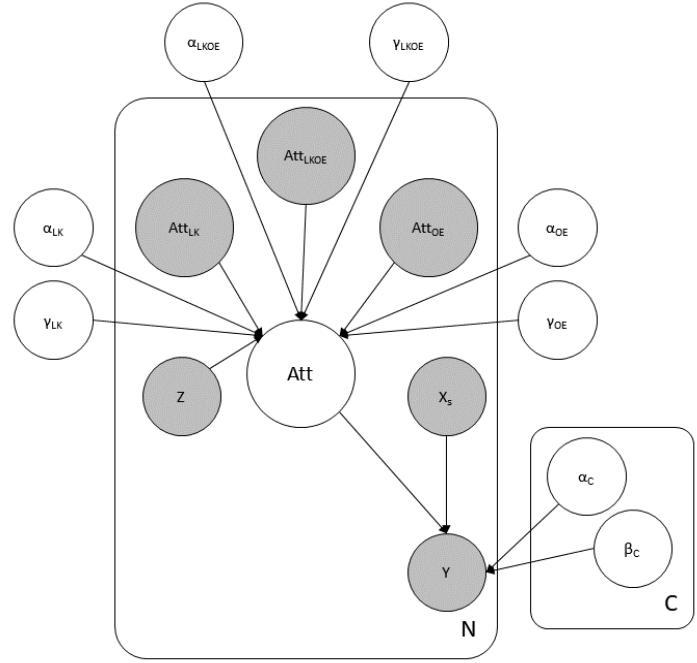

**Fig. 5.** PGM for the proposed model

Similar to the individual models, we assume the latent attitudes to be identically and independently distributed ("i.i.d") [K-dimensional] with a mean estimated using equation (23) and unit standard deviation. We assume a normal distribution (mean: 0; std. dev: 1) for the constants ($\alpha_{lk}$, $\alpha_{lkoe}$, and $\alpha_{oe}$) and coefficients ($\gamma_{lk}$, $\gamma_{lkoe}$, and $\gamma_{oe}$). We assume an approach similar to that for the individual models, with a similar estimation approach and distribution for the variables for the mode choice variable.

**Generative Process**

1. For equation $k \in \{1, …, K\}$:
   a. *Draw coef* $\alpha_{lk\,(k)} \sim Normal\left(\alpha_{lk\,(k)} \mid 0, 1\right)$ (eq. 17)
   b. *Draw coef* $\gamma_{lk\,(k)} \sim Normal\left(\gamma_{lk\,(k)} \mid 0, 1\right)$ (eq. 18)
   c. *Draw coef* $\alpha_{lkoe\,(k)} \sim Normal\left(\alpha_{lkoe\,(k)} \mid 0, 1\right)$ (eq. 19)
   d. *Draw coef* $\gamma_{lkoe(k)} \sim Normal\left(\gamma_{lkoe\,(k)} \mid 0, 1\right)$ (eq. 20)
   e. *Draw coef* $\alpha_{oe\,(k)} \sim Normal\left(\alpha_{oe\,(k)} \mid 0, 1\right)$ (eq. 21)
   f. *Draw coef* $\gamma_{oe\,(k)} \sim Normal\left(\gamma_{oe\,(k)} \mid 0, 1\right)$ (eq. 22)
   g. *Draw coef* $Att_{(k)} \sim Normal\left(Att_{(k)} \mid \begin{array}{l}(Z == 1) * (\alpha_{lk\,(k)} + \gamma_{lk\,(k)} * Att_{lk\,(n)}) + \\ (Z == 2) * (\alpha_{lkoe\,(k)} + \gamma_{lkoe\,(k)} * Att_{lkoe\,(n)}) + \\ (Z == 3) * (\alpha_{oe\,(k)} + \gamma_{oe\,(k)} * Att_{oe\,(n)})\end{array}, 1\right)$ (eq. 23)

2. For each class $c \in \{1, \ldots, C\}$:
   a. *Draw coef* $\alpha_c \sim Normal\,(\alpha_c \mid 0, 1)$ (eq. 24)
   b. *Draw coef* $\boldsymbol{\beta_c} \sim Normal\,(\boldsymbol{\beta_c} \mid 0, 1)$ (eq. 25)
   c. *Draw choice* $\boldsymbol{Y_c} \sim Multinomial\,(\boldsymbol{Y_c} \mid Softmax\,([\boldsymbol{X_s}, \boldsymbol{A_t}], \alpha_1 \ldots \alpha_C, \boldsymbol{\beta_1} \ldots \boldsymbol{\beta_C}))$ (eq. 26)

**Joint Probability Distribution**

$$p(\alpha_{LK}, \gamma_{LK}, \alpha_{LKOE}, \gamma_{LKOE}, \alpha_{OE}, \gamma_{OE}, \alpha_C, \beta_C, Att_{LK}, Att_{LKOE}, Att_{OE}, \boldsymbol{Att}, Z, \boldsymbol{X_S}, Y)$$

$$= p(\alpha_{LK})p(\gamma_{LK})p(\alpha_{LKOE})p(\gamma_{LKOE})p(\alpha_{OE})p(\gamma_{OE}) \left(\prod_{c=1}^{C} p(\alpha_c)\right)\left(\prod_{c=1}^{C} p(\beta_c)\right)$$

$$\left(\prod_{n=1}^{N} \begin{array}{c} p(Att_{LK})\,p(Att_{LKOE})\,p(Att_{OE})\,p(Z) \\ p(\boldsymbol{Att} \mid Att_{LK}, Att_{LKOE}, Att_{OE}, Z, \alpha_{LK}, \gamma_{LK}, \alpha_{LKOE}, \gamma_{LKOE}, \alpha_{OE}, \gamma_{OE}) \\ p(\boldsymbol{X_S})\,p(Y \mid \boldsymbol{Att}, \boldsymbol{X_S}, \alpha_C, \beta_C) \end{array}\right)$$ (eq. 27)

*4. 4. Predicting attitudes in the counterfactual instrument*

Having estimated the models using our proposed framework, we can now use the estimated coefficients to evaluate the corresponding scores using the counterfactual instrument, i.e., if one had an open-ended response, what would the corresponding Likert scale response be? And having a Likert scale response, what would the open-ended response topics look like? We achieve this by using the coefficients and the values for the latent attitudes. A modified version of the Gibbs Sampling is used to accomplish this and is illustrated in Fig. 6 (Gelfand, 2000).

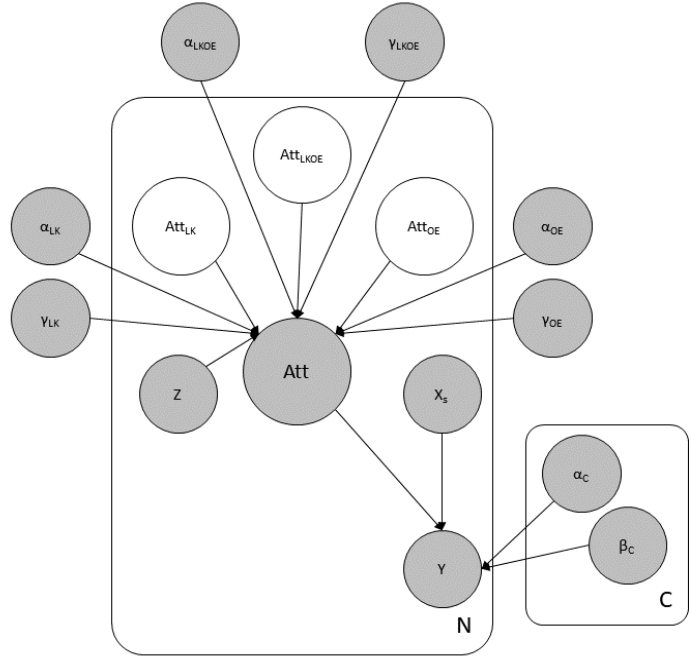

**Fig. 6.** Modified framework to map responses (note that now all previously estimated parameters are shaded, i.e. they are "observed")

The latent attitudes ($y_{att}$) are a multidimensional function, and so are the coefficients $\alpha$ and $\gamma$. Assuming X to be the response to a Likert scale response or the topic proportions for an open-ended response, equations for the latent attitudes can be written as: -

$$y_{att} = \alpha + \gamma_1 X_1 + \gamma_2 X_2 + \cdots + \gamma_n X_n$$ (eq. 28)

However, our research presents additional challenges to adopting the conventional approach, as the sets of variables ($X_1, X_2, …, X_n$) are not mutually independent. Each of these variables ($X_1$) are vectors that together must sum up to one. We use vectorization in Linear Algebra and draw the variables from a Dirichlet distribution to account for this.

**5. Exploratory analysis on attitudinal variables**

*5. 1. Treatment of closed-ended responses*

To assess the internal reliability of the Likert scale responses, we computed the Cronbach's Alpha values (presented in Table 1) – which were high for both versions of the questionnaire. As compared to Ver_LK, slightly higher values were observed for Ver_LKOE of the questionnaire. To test the validity of the questionnaire, we compared the average scores for the Likert scale responses in two groups (aware of AVs v/s unaware of AVs), and the respondents aware of AVs seem to have responded correctly to the statements. The differences between the two groups were statistically significant (t-stats) for both versions ("Ver_Lk" – 2.504 and "Ver_LKOE" – 2.638), which would support the validity of the questionnaire. We did not report the p-values and degrees of freedom because the t-stats were higher than the cut-off values.

*5. 2. Extraction of information from open-ended responses*

*5. 2. 1 Data cleaning*

Before using Topic Modeling methods, one needs to pre-process textual data[6]. We started by using "Grammarly" (Grammarly, 2021) to automatically correct mistakes (spelling and grammatical errors). Also, for phrases involving negation, such as "don't trust" or "do not trust", we applied "Regular Expressions" or regex (Python, 2021) to replace them with prefixed expressions (e.g. "no_trust"). We used regex to capture synonyms and abbreviations and unify number representations (e.g., 3 becomes "three"). Doing this is also a step towards dimensionality reduction – which we believe is critical in our analysis to reduce model complexity and better generalize.

We then applied "tokenization" of the text, splitting the text into words while removing punctuations. The subsequent stage is reducing words to their root form, called lemmatization (e.g., "travelled", "travel", "travelling" becomes "travel") and stemming (e.g., "easiness", "easy" becomes "easi")[7]. After that, we removed words that have no relevant semantic meaning for our task, named "stopwords", such as pronouns, prepositions, common verbs (e.g., "the", "of", "they", "I", "is", "are"). Words that are redundant for the study context should also be added to the stopword list. In this particular case, words such as "Autonomous", "Vehicles", "Cars", etc. are considered redundant. For several of these data preparation treatments, we used the Python NLTK package[8].

This data cleaning process is iterative and time-consuming. At any given stage, the analyst needs to decide the appropriate degree of data cleaning by examining the topics for words that are redundant, irrelevant, or others that can be merged into one concept (e.g., synonyms). The goal is to obtain topics with meaningful interpretation and statistical significance. In the case at hand, after this data preparation process, we obtained a significant reduction in the average number of words per open-ended response (~65%) – from 22.03 words to 7.78 words for PU and 20.50 to 8.07 for PEoU.

---

[6] https://towardsdatascience.com/nlp-extracting-the-main-topics-from-your-dataset-using-lda-in-minutes-21486f5aa925
[7] Hereinafter, in our explanations, we refer to the "stemmed" version of words.
[8] https://www.nltk.org/

*5. 2. 2 Results from Topic Models*

The top five words from each of the extracted topics are presented in Table 5. Having estimated the topics, we searched the words in a given topic in the survey responses. Based on the information conveyed by the respondents, we labelled the topics. To refer to the topics, we use "Top", followed by the number of open-ended questions and the topic number. For instance, Top11 refers to the first topic extracted for the first open-ended question.

**Table 5**

Mapping between Closed- and Open-ended Responses

| Psychological Construct | Characteristics | Likert Scale Question | Extracted Topics (top 5 words) |
|---|---|---|---|
| Perceived Ease of Use | Easy for me | AV_LeaEa- Learning to use Autonomous Vehicles will be easy for me | Top11- use, easi, technolog, work, get |
| | Get them to do what I want them to do | AV_WrkEa- I will find it easy to get Autonomous Vehicles to do what I want them to do | Top12- drive, road, human, mani, accid |
| | Easy to become skillful | AV_SklEa- It will be easy for me to become skillful at using Autonomous Vehicles | Top13- drive, control, driver, make, easier |
| | Easy to use | AV_UseEa- I will find Autonomous Vehicles easy to use | Top14- oper, go, everyth, assum, user |
| Perceived Usefulness | Useful in meeting travel needs | AV_TrNed- Using Autonomous Vehicles will be useful in meeting my travel needs | |
| | Perform other tasks | AV_OthAc- Autonomous Vehicles will let us do other tasks such as eating, watching a movie, be on a cell phone during my trip | Top22- drive, driver, thing, work, make  Top26- take, go, attent, pay, use |
| | Decrease my accident risk | AV_DeAcc- Using Autonomous Vehicles will decrease my accident risk | Top27- driver, help, transport, safety, safer |
| | Relieve my stress of driving | AV_ReStr- Using Autonomous Vehicles will relieve my stress of driving | Top25- use, drive, need, technolog, situat |
| | Useful when I'm impaired | AV_UsImp- I find Autonomous Vehicles to be useful when I'm impaired | |
| | Mobility for the disabled | | Top24- drive, use, get, help, disabl |
| | Lesser congestion | | Top23- accid, human, traffic, reduc, help |
| | Better for environment | | Top21- time, better, environ, make, save |

| | |
|---|---|
| Saves time | Top21- time, better, environ, make, save |

We extracted four topics from the open-ended question for PEoU, and the first extracted topic (Top11) was primarily about the ease of getting it to work, learn and gain trust. The second topic (Top12) emphasized the need for human presence to respond to uncertainties, which could otherwise lead to errors. Ease of operation was covered in the third topic (Top13), while in the fourth topic (Top14), they also discussed the additional benefits brought about by self-navigation. Seven broad themes were extracted from the responses to the perceived usefulness of AVs. Respondents believed that AVs might save on travel time and make travel more environmentally friendly (Top21). On the ability to work during the trip, participants shared contrasting views. Respondents believed that AVs might facilitate working during travel (Top22); this may require additional attention, which could negatively affect their work (Top26). AVs could make travelling safer and mitigate congestion (Top23), make parking easier (Top27) and ensure mobility for the disabled (Top24). Many participants emphasized the need for human control while using AVs (Top25).

*5. 3. Extracted topics versus Likert scale topics*

After extracting information from open-ended responses, we analyzed the coherence of the answers from closed- and open-ended questions. The closed- and open-ended questions were carefully designed to ensure coherence in our surveys. With reference to Table 5, it is encouraging to note that a related topic was extracted from the open-ended response for most of the individual statements presented in the closed-ended questions. Furthermore, the open-ended responses allowed for the extraction of more information on the topic. For questions related to the "Perceived Ease of Use", it was possible to achieve a one-to-one mapping between the closed- and open-ended responses. In addition to the four aspects of the closed-ended questions, Topic Models also highlighted the need for human control. Answers to the question on the "Perceived Usefulness" were not directly mapped to the Likert scale questions "Useful in meeting travel needs" and "Useful when I'm impaired". Having identified the remaining characteristics presented in the Likert scale questions, Topic Models identified other aspects such as "mobility for the disabled", "congestion reduction", "environmental friendliness", and "travel time reduction" that makes AVs worthwhile.

## 6. Estimation results and salient findings

Having extracted information from the closed- and open-ended responses, we proceeded with estimating the models for the intention to use AVs for commute trips. Variables depicting attitudes, socio-demographic and travel characteristics of the individuals, familiarity with AVs and SP experiment characteristics were used as explanatory variables. The models were estimated using the frameworks proposed earlier in Fig. 4 and Fig. 5. The models were estimated using Pyro (Bingham et al., 2019) in a GPU. Pyro is a universal probabilistic programming language written in Python, with PyTorch supports on the backend. In order to draw the Bayesian Inference, we used the Stochastic Variational Inference proposed by Hoffman et al. (2013).

In order to benchmark the performance of the models, we estimated three models separately for each of the datasets (named "Ind") and compared the performance of the proposed model (called "Prop" – a combined model for the three datasets) with the performance of each of these individual models. We performed this analysis using both a training set (80%) and a test set (20%) of randomly selected responses. The performance measures of the individual models and proposed models' performance measures are presented in Table 6. As discussed earlier, our proposed model estimates a single model for the three datasets. We then used the model estimates to predict the values for each of these datasets (Ver_LK, Ver_LKOE, and Ver_OE), and these predicted values are used to compute the performance measures. These results for each dataset are presented in the columns named "Prop". For example, in the second and third columns, we show the values for the individual and proposed models for Ver_LK.

We compared the performance using the initial log-likelihood ($LL_I$), log-likelihood with respect to constants ($LL_C$), final log-likelihood ($LL_F$) and the McFadden $\rho^2$ value. The McFadden $\rho^2$ value is computed with respect to constants ($\rho_c^2$), so as to account for the improvement of the model with respect to constants, and the adjusted McFadden $\rho^2$ is used to account for the improvements with the estimation after considering the estimated parameters. It should be noted that while computing the adjusted values for $\rho^2$ and Count $R^2$, we did not account for the parameters of the LDA model itself. In addition to this, we present the goodness-of-fit measures such as the Count $R^2$ (Group, 2021) and the $F_1$ scores (van Rijsbergen, 1979). In Appendix A, we present a short description of these performance measures along with the formula for its computation.

This analysis consists of two parts – so as to quantify the improvements with a. *the introduction of open-ended questions before the set of closed-ended responses;* b. *the use of open-ended questions*. In pursuing the first part of the analysis, we revisit and investigate further the findings from our previous research (Baburajan et al., 2020). The second part of the analysis extends said research further by relying only on open-ended questions to measure attitudes.

**Table 6**

Goodness-of-fit Measures for Training and Test Set

|  |  | Ver_LK | | Ver_LKOE | | Ver_OE | |
|---|---|---|---|---|---|---|---|
|  |  | Ind | Prop | Ind | Prop | Ind | Prop |
| Training set | $LL_I$ | -5328.27 |  | -5211.82 |  | -5127.22 |  |
|  | $LL_C$ | -5021.83 | -5059.77 | -5106.44 | -5111.87 | -5080.24 | -5100.86 |
|  | $LL_F$ | -4402.87 | -4451.77 | -4552.77 | -4477.36 | -4654.19 | -4733.33 |
|  | $\rho^2$ | 0.17 | 0.16 | 0.13 | 0.14 | 0.09 | 0.08 |
|  | $\rho_c^2$ | 0.12 | 0.12 | 0.11 | 0.12 | 0.08 | 0.07 |
|  | K | 97 |  | 97 |  | 63 |  |
|  | Adj. $\rho^2$ | 0.16 | 0.15 | 0.11 | 0.12 | 0.08 | 0.06 |
|  | Count $R^2$ | 59.79 | 58.50 | 59.95 | 58.26 | 50.74 | 49.18 |
|  | Adj. Count $R^2$ | 35.09 | 35.87 | 43.49 | 44.02 | 33.52 | 32.94 |
|  | $F_1$ Score | 0.54 | 0.53 | 0.57 | 0.56 | 0.49 | 0.48 |
|  | Accuracy | 0.60 | 0.58 | 0.60 | 0.58 | 0.51 | 0.49 |
|  | Precision | 0.55 | 0.54 | 0.58 | 0.57 | 0.50 | 0.49 |
|  | Recall | 0.52 | 0.52 | 0.56 | 0.55 | 0.49 | 0.47 |
| Test set | K | 97 |  | 97 |  | 63 |  |
|  | Adj. $\rho^2$ | -0.10 | -0.05 | -0.10 | 0.06 | -0.04 | -0.01 |
|  | Count $R^2$ | 54.58 | 53.36 | 55.74 | 55.90 | 46.54 | 46.64 |
|  | Adj. Count $R^2$ | 24.60 | 26.31 | 38.94 | 40.17 | 31.17 | 32.90 |

| | | | | | | |
|---|---|---|---|---|---|---|
| F$_1$ Score | 0.45 | 0.44 | 0.53 | 0.53 | 0.45 | 0.45 |
| Accuracy | 0.55 | 0.53 | 0.56 | 0.56 | 0.47 | 0.47 |
| Precision | 0.45 | 0.43 | 0.53 | 0.54 | 0.45 | 0.45 |
| Recall | 0.45 | 0.45 | 0.52 | 0.53 | 0.45 | 0.45 |

For the first part of the analysis, we compared the performance of the models estimated using Ver_LK and Ver_LKOE of the questionnaire. To begin with, we included all attitudinal variables (PU, PEoU, PSR, PPR, Tr and Att) to predict the intention to use Autonomous Vehicles as a mode for commute trips. We compared the performance of these models with the attitudinal variables in the TAM model proposed by Davis, Jr (1985). Compared to the latter, we did not see any improvement (accuracies for Ver_LK, Ver_LKOE and Ver_OE are 0.57, 0.54 and 0.44, respectively), particularly after accounting for the additional parameters. Hence, we limit our analysis and discussion to the latter (full results/data are available upon request).

In terms of the results of Table 6, improvements in terms of $\rho^2$ and count $R^2$ were highest for models estimated using Ver_LK of the questionnaire. However, after accounting for the constants, the models estimated using Ver_LKOE showed performance improvement and performed best in the Count $R^2$ count. This difference in terms of performance is reasonable as, unlike other datasets, Ver_LK had a higher proportion of respondents choosing "Regular Cars", which was significantly higher than the other options presented to the respondents. Having almost 50% of the respondents selecting a given alternative is likely to influence that models' forecasting capability substantially. Exploring the second aspect of this analysis, we could not identify any improvements in performance using questionnaires that used only open-ended questions to measure attitudes.

Now analyzing results on the test set, we compared the F$_1$ scores, and the highest values were obtained for Ver_LKOE of the questionnaire. Reasonable and comparable values were obtained for Ver_LK and Ver_OE questionnaires. Similar trends are observed for the test set; however, in certain cases, the performance of goodness-of-fit measures for Ver_LK was lower than for Ver_OE. Interestingly, in the test set, the proposed model for Ver_LKOE performed significantly better than Ver_LK and Ver_OE. This becomes even more pronounced after accounting for the estimated parameters (Adj. Count $R^2$), thus confirming our assumption that about 50% of the respondents choosing an alternative could be the reason for the models' better performance for Ver_LK.

When evaluating the performance of our proposed model with the individual models, it is worth noting that the proposed model performs better when compared to the individual models. For the training set, this is observed only for Ver_LKOE. However, it performs significantly better for all three versions of the questionnaires for the test set. This is true for the various goodness-of-fit measures, such as the adjusted Count $R^2$ and the F$_1$ scores.

The primary objective of this paper is to present the proposed framework and demonstrate its capability to predict responses for a given questionnaire type using the coefficients and attitudes estimated based on responses to another questionnaire type. Accordingly, we do not present an in-depth analysis of the estimated coefficients' nature, magnitude, and statistical significance. The estimated coefficients are presented in Appendix B. We would, however, highlight that the effects of the estimated coefficients align with those results observed by other researchers. As the Stated Preference section of our experimental design was adopted from Haboucha et al. (2017), it is reassuring to note that our findings resonate with theirs.

## 7. Mapping of responses from other questionnaires

Having estimated the coefficients and the latent attitudes, we used the framework (Fig. 6) to predict responses for other questionnaire versions, i.e., the counterfactual instrument. In other words, for the questionnaires with Likert scale responses (and the inferred latent attitudinal response), we obtain the corresponding topic distributions. We obtain the corresponding Likert scale responses for the questionnaires with open-ended responses. We demonstrate these capabilities by presenting the results in Table 8, Table 9, Table 10 and Fig. 7. The reader should note that the mean values are merely for understanding the capabilities. To make predictions for a given dataset, one should use the estimated latent attitudes for those individuals and compute the values for the explanatory variables.

We present the results of the mapping (the average of the generated values) of the Likert scale and open-ended responses in Table 8 for Ver_LK and in Table 9 for Ver_LKOE of the questionnaire, respectively. Abbreviations have been used for the various levels of the Likert scale responses (SD – Strongly Disagree, Disag – Disagree, Neut – Neutral, Agree – Agree and SA – Strongly Agree). Table 10 presents the mapping of the observed topic proportions for open-ended questions and the averages generated for Likert scale responses for Ver_LK and Ver_LKOE. The abbreviations for variables related to closed-ended questions and topics are the same as in Table 5. In relation to the columns in Table 8, for the Likert scale responses, one could directly map the observed and the generated averages for Ver_LK and Ver_LKOE of the questionnaire. However, for the topic proportions of the open-ended questions, readers should not expect a direct correspondence between Likert scale questions and topics in the same row (for instance, there is no direct correspondence between AV_LeaEa and Top11 or between AV_WrkEa and Top12).

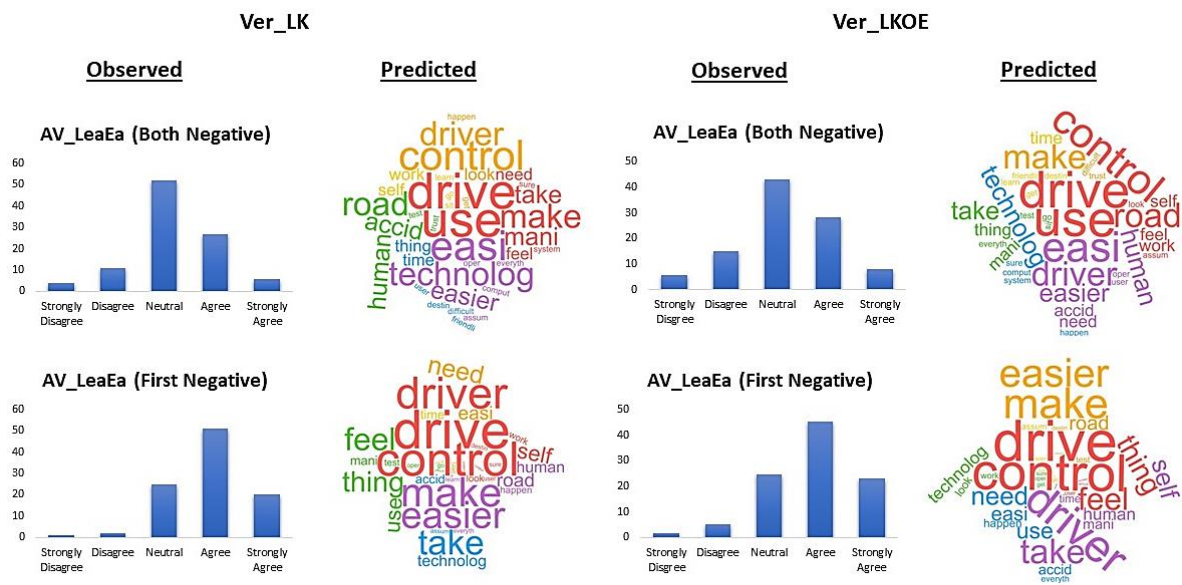

**Fig. 7.** Predictions using the proposed framework

The proposed approach allows respondents to choose the questionnaire based on type of interest. If the existing models used by researchers/analysts utilize responses from Likert scale (or open-ended) questions that the respondents have answered using open-ended (or Likert scale) responses, using this framework will enable them to deduce approximate Likert scale (or open-ended) responses which could then be used to predict behavior. To demonstrate this to our readers, we present in Fig. 7 the average values for the topic proportions for the two closed-ended questionnaires (Ver_LK and Ver_LKOE). Readers should note that this is merely to demonstrate the values, and for practical purposes, one should use the predictions for each question for each respondent. In our study, attitudes are two dimensional, and we segment respondents into two categories ("Both Negative" and "First Negative") using the nature (sign) of the estimated attitudes. For each dataset (Ver_LK and Ver_LKOE), we estimate the topic proportions for the observed scale responses and obtain the word clouds.

We will now demonstrate how practitioners can use our proposed framework. For reasons of simplicity, we will obtain the aggregate shares for each of the different modes by computing the utility for an average respondent. We assume that the state-of-the-art models use closed-ended responses to measure attitudes before integrating them into the mode choice models. This means that practitioners cannot calculate the aggregate shares using the state-of-the-art models to measure attitudes using an open-ended question. They have to estimate separate models. In such circumstances, analysts can use our proposed method, as we demonstrate below. By using eq. 23, we can compute the scores for attitudes for Ver_LK (-23.49, 44.04), Ver_LKOE (-24.23, 9.99), and Ver_OE (-25.29, 44.04). The analyst can then compute the shares of the different modes for the other datasets (Ver_LK, Ver_LKOE, and Ver_OE) using eq. 27. We compute the aggregate shares for the population by assuming the responses of an average person – using average values for all explanatory variables except the latent attitudes. It is assumed that analysts use models based on closed-ended responses. Hence, Ver_LK is observed in this example, and we need to generate values for Ver_LKOE and Ver_OE. The attitudes can be computed for: -

- Ver_LK using the observed values for attitudinal variables from Table 8 (columns 2-6)
- Ver_LKOE using the generated values for attitudinal variables from Table 9 (columns 7-11)
- Ver_OE using the generated values for attitudinal variables from Table 10 (columns 9-13)

The shares computed using the utility equation estimated using Ver_Lk are presented below in Table 7.

**Table 7**

Predicted Shares using utility equation estimated using Ver_LK

| Dataset used | Regular Car | Private AV | Shared AV |
| --- | --- | --- | --- |
| Ver_LK | 56.45 | 29.87 | 13.68 |
| Ver_LKOE | 69.99 | 19.16 | 10.84 |
| Ver_OE | 53.59 | 32.55 | 13.86 |

**Table 8**

Mapping of the Likert Scale Responses for Ver_LKOE and Ver_OE (Attitudes estimated for Ver_LK)

|  | SD | Disag | Neut | Agree | SA | SD | Disag | Neut | Agree | SA | Top # | Top Prop (%) | Word_1 | Word_2 | Word_3 | Word_4 | Word_5 |
|---|---|---|---|---|---|---|---|---|---|---|---|---|---|---|---|---|---|
|  | Ver_LK (Observed Averages) | | | | | Ver_LKOE (Generated Averages) | | | | | Ver_OE (Generated Topic Proportions) | | | | | | |
| AV_LeaEa | 5.79 | 10.04 | 35.68 | 35.67 | 12.82 | 10.63 | 23.51 | 7.71 | 15.00 | 26.13 | Top11 | 13.95 | use | easi | technolog | work | get |
| AV_WrkEa | 5.27 | 13.07 | 43.59 | 30.30 | 7.77 | 10.94 | 16.62 | 16.90 | 10.85 | 27.66 | Top12 | 27.47 | drive | road | human | mani | accid |
| AV_SklEa | 4.60 | 10.07 | 32.08 | 41.59 | 11.66 | 9.28 | 14.13 | 12.91 | 12.16 | 34.50 | Top13 | 49.98 | drive | control | driver | make | easier |
| AV_UseEa | 4.40 | 10.32 | 36.27 | 39.22 | 9.79 | 12.49 | 8.51 | 11.68 | 36.37 | 13.92 | Top14 | 9.91 | oper | go | everyth | assum | user |
| AV_TrNed | 7.00 | 12.04 | 33.63 | 34.23 | 13.09 | 6.69 | 29.73 | 6.38 | 22.89 | 17.28 | Top21 | 12.44 | time | better | environ | make | save |
| AV_OthAc | 11.57 | 15.57 | 26.08 | 34.87 | 11.91 | 42.92 | 13.30 | 7.44 | 8.14 | 11.18 | Top22 | 21.35 | drive | driver | thing | work | make |
| AV_DeAcc | 12.42 | 22.34 | 37.29 | 21.56 | 06.39 | 9.93 | 21.87 | 16.37 | 21.41 | 13.39 | Top23 | 17.63 | accid | human | traffic | reduc | help |
| AV_ReStr | 12.74 | 22.66 | 27.70 | 27.48 | 9.42 | 13.69 | 12.13 | 8.30 | 27.55 | 21.31 | Top24 | 3.53 | drive | use | get | help | disabl |
| AV_UsImp | 11.34 | 12.87 | 26.56 | 35.82 | 13.41 | 37.79 | 2.51 | 12.51 | 17.15 | 13.02 | Top25 | 22.78 | use | drive | need | technolog | situat |
|  |  |  |  |  |  |  |  |  |  |  | Top26 | 15.38 | take | go | attent | pay | use |
|  |  |  |  |  |  |  |  |  |  |  | Top27 | 6.89 | driver | help | transport | safeti | safer |

Note: For the acronyms used in the table, kindly refer to Table 5

**Table 9**

Mapping of the Likert Scale Responses for Ver_LK and Ver_LKOE (Attitudes estimated for Ver_LKOE)

|  | SD | Disag | Neut | Agree | SA | SD | Disag | Neut | Agree | SA | Top # | Top Prop (%) | Word_1 | Word_2 | Word_3 | Word_4 | Word_5 |
|---|---|---|---|---|---|---|---|---|---|---|---|---|---|---|---|---|---|
|  | Ver_LKOE (Observed Averages) | | | | | Ver_LK (Generated Averages) | | | | | Ver_OE (Generated Topic Proportions) | | | | | | |
| AV_LeaEa | 6.16 | 10.01 | 29.63 | 37.16 | 17.03 | 12.69 | 16.38 | 16.10 | 18.80 | 16.04 | Top11 | 12.34 | use | easi | technolog | work | get |
| AV_WrkEa | 6.44 | 12.91 | 40.49 | 31.52 | 8.64 | 8.30 | 10.17 | 7.44 | 32.63 | 21.46 | Top12 | 24.90 | drive | road | human | mani | accid |
| AV_SklEa | 5.56 | 8.24 | 27.23 | 42.34 | 16.62 | 7.40 | 12.14 | 22.05 | 21.06 | 17.34 | Top13 | 52.40 | drive | control | driver | make | easier |
| AV_UseEa | 5.74 | 10.58 | 29.83 | 39.70 | 14.15 | 31.25 | 16.03 | 9.56 | 5.17 | 17.99 | Top14 | 10.36 | oper | go | everyth | assum | user |
| AV_TrNed | 10.20 | 11.77 | 25.67 | 33.87 | 18.49 | 16.78 | 36.30 | 5.99 | 9.92 | 11.00 | Top21 | 11.89 | time | better | environ | make | save |
| AV_OthAc | 11.12 | 13.64 | 21.60 | 33.72 | 19.91 | 19.46 | 17.57 | 16.94 | 10.18 | 15.85 | Top22 | 21.50 | drive | driver | thing | work | make |
| AV_DeAcc | 14.47 | 17.13 | 29.96 | 25.54 | 12.90 | 24.32 | 8.91 | 14.99 | 15.07 | 16.71 | Top23 | 18.40 | accid | human | traffic | reduc | help |
| AV_ReStr | 14.53 | 18.34 | 22.15 | 31.02 | 13.95 | 10.32 | 32.76 | 4.55 | 18.10 | 14.27 | Top24 | 3.63 | drive | use | get | help | disabl |

| | | | | | | | | | | | Top25 | 21.12 | use | drive | need | technolog | situat |
|---|---|---|---|---|---|---|---|---|---|---|---|---|---|---|---|---|---|
| AV_UsImp | 12.13 | 12.58 | 19.50 | 32.96 | 22.83 | 9.84 | 3.84 | 41.61 | 7.99 | 16.72 | Top26 | 16.84 | take | go | attent | pay | use |
| | | | | | | | | | | | Top27 | 6.62 | driver | help | transport | safeti | safer |

Note: For the acronyms used in the table, kindly refer to Table 5

**Table 10**

Mapping of the Likert Scale Responses for the Extracted Topics from Open-ended Responses (Attitudes estimated for Ver_OE)

| Top # | Top Prop (%) | Word_1 | Word_2 | Word_3 | Word_4 | Word_5 | | SD | Disag | Neut | Agree | SA | SD | Disag | Neut | Agree | SA |
|---|---|---|---|---|---|---|---|---|---|---|---|---|---|---|---|---|---|
| | Ver_OE (Observed Topic Proportions) | | | | | | | Ver_LK (Generated Averages) | | | | | Ver_LKOE (Generated Averages) | | | | |
| Top11 | 9.68 | use | easi | technolog | work | get | AV_LeaEa | 13.28 | 18.58 | 16.01 | 19.05 | 16.05 | 10.27 | 23.93 | 7.74 | 14.52 | 26.52 |
| Top12 | 41.98 | drive | road | human | mani | accid | AV_WrkEa | 8.69 | 12.00 | 7.27 | 34.52 | 20.49 | 10.94 | 16.66 | 16.58 | 11.37 | 27.42 |
| Top13 | 37.60 | drive | control | driver | make | easier | AV_SklEa | 9.15 | 14.84 | 21.40 | 20.44 | 17.15 | 9.20 | 13.61 | 13.22 | 11.96 | 35.00 |
| Top14 | 10.74 | oper | go | everyth | assum | user | AV_UseEa | 30.69 | 17.76 | 9.52 | 5.39 | 19.62 | 12.44 | 8.51 | 11.70 | 36.17 | 14.17 |
| Top21 | 3.81 | time | better | environ | make | save | AV_TrNed | 16.49 | 37.79 | 6.79 | 10.63 | 11.28 | 6.85 | 29.86 | 6.20 | 22.07 | 18.01 |
| Top22 | 15.70 | drive | driver | thing | work | make | AV_OthAc | 21.36 | 19.38 | 16.80 | 9.80 | 15.65 | 40.71 | 13.89 | 7.60 | 8.72 | 12.06 |
| Top23 | 21.59 | accid | human | traffic | reduc | help | AV_DeAcc | 26.95 | 8.85 | 16.28 | 14.74 | 16.15 | 9.94 | 22.12 | 16.34 | 21.79 | 12.79 |
| Top24 | 8.57 | drive | use | get | help | disabl | AV_ReStr | 9.84 | 34.06 | 5.36 | 17.39 | 16.32 | 13.60 | 12.28 | 8.33 | 28.00 | 20.78 |
| Top25 | 12.50 | use | drive | need | technolog | situat | AV_UsImp | 10.10 | 4.35 | 42.96 | 8.26 | 17.31 | 36.60 | 2.75 | 12.13 | 16.79 | 14.70 |
| Top26 | 9.31 | take | go | attent | Pay | use | | | | | | | | | | | |
| Top27 | 28.52 | driver | help | transport | safeti | safer | | | | | | | | | | | |

Note: For the acronyms used in the table, kindly refer to Table 5

## 8. Conclusions

Using a dataset from the US collected between January and March 2020, we estimated models that predict the intention to use AVs as a mode for commute trips. We included respondents' socio-demographic and travel characteristics, familiarity with AVs, attitudes towards AVs and variables from the Stated Preference survey as explanatory variables. In measuring attitudes, we used three different questionnaires. The collected data was representative of the US population based on gender, ethnicity, and regional diversity.

As observed in our previous study (Baburajan et al., 2020), we find Topic Modeling effective for extracting information from open-ended responses. We observed correspondence between the extracted topics and the Likert scale questions. This is particularly interesting to the researchers using open-ended questions because it significantly reduces the effort to extract information from open-ended responses. In addition to this, Topic Models can be very effective in getting topics out of respondents that were unforeseen from the survey design perspective. This is particularly relevant in analyzing attitudes in relatively new areas where researchers may have limited information. Furthermore, extracting information from open-ended responses is likely to be affected by the subjective bias of different coders (Glazier et al., 2021). Analysts have been circumventing this problem by developing coding strategies independently before arriving at coding that is considered reliable. Using the Topic Modeling approach assists researchers in overcoming this issue, as the algorithm is indifferent to individual biases, can handle very large datasets, and reveals statistically significant (collections of) concepts that can objectively support coding.

However, its use warrants due diligence as researchers require expert judgment in the often time-consuming data cleaning process to avoid removing important information from the responses. Furthermore, based on our experience, each question may demand the use of "stopwords" that are question-specific and probably data-specific, which should be the object of further research. This is also the case with combining words. Moreover, the languages supported in NLTK are limited (23), which could pose difficulties for extracting information from open-ended responses for surveys conducted in languages not supported by NLTK.

Our first objective was to develop a framework that allows respondents to choose their preferred questionnaire type to generate responses to attitudinal questions. We developed this framework using Probabilistic Graphical Models, which we later use to generate responses for an explanatory variable of interest. Our proposed framework provides flexibility to both analysts and participants. For instance, from the analysts' perspective, the established models are estimated using Likert scale responses. However, respondents may prefer to respond to open-ended questions; our proposed framework will thus allow analysts to extract corresponding Likert scale responses for the individuals' open-ended responses. The proposed framework could be beneficial when respondents are willing to express their attitudes using voice-based open-ended surveys, thus significantly reducing the burden involved.

When comparing the performance of our proposed model with the individual models, it is encouraging to note that it performs as good as the individual models, which were also validated using a test set (20%). In line with the results from our previous study (Baburajan et al., 2020), the Likert scale questions with the open-ended questions before them (Ver_LKOE) performed better than Ver_LK, particularly for the test set. However, the models estimated using open-ended responses (Ver_OE) did not perform as good as the Likert scale questions. This is interesting from the perspective of an analyst/researcher as one need not opt for open-ended questions but could use Likert scale questions with open-ended questions before them. A plausible explanation for the poor performance of the models estimated using open-ended responses may have to do with data cleaning.

Future avenues for research include anchoring information extraction from open-ended responses based on various linguistic theories. This is particularly relevant in light of the findings of other researchers regarding the use of Artificial Intelligence, which echo the aphorism "Garbage in Garbage out" (Henman, 2020). More specifically, in extracting information from open-ended responses, it may be of interest to use other Natural Language Processing techniques, such as word embeddings (Jurafsky and Martin, 2009), to extract information and compare performance with LDA. With a view to improving open-ended surveys, researchers should carry out in-depth interviews that might help contextualize appropriate open-ended questions, improve the quality of responses and devise strategies for data cleaning. In addition to this, researchers discuss the potential hypothetical bias in Stated Preference experiments, particularly in studies about future technologies. Although there has been debate on the prevalence and importance of analyzing hypothetical bias, it is an area that still requires considerable attention, particularly in transportation studies that involve large public investments (Haghani et al., 2021a). Haghani et al. (2021b) discuss the different methods researchers use to mitigate hypothetical bias. We believe this might be more relevant for online surveys, particularly voice-based surveys. Therefore, analyzing the tone of responses (interested or disinterested in the survey) and the speed with which respondents answer voice-based surveys (either revealing haste to complete the survey or thinking carefully before responding) might provide meaningful insights into this issue.

## Acknowledgements

Funding: This work was supported by the Fundação para a Ciência e a Tecnologia (FCT, IP), Portugal, [grant number PD/BD/129394/2017].

We would like to thank two anonymous reviewers for carefully reading our manuscript and offering insightful comments and suggestions. The authors would also like to acknowledge Dr Francisco Antunes (Post-Doctoral Researcher, Technical University of Denmark), Mr Jithin P. Zachariah (Research Scholar, Indian Institute of Technology, Roorkee), Mrs Elna James Kattoor (Assistant Instructor, University of Toronto) for their comments and suggestions during the process of revising this paper.

# Appendix A

**Table 11**
Goodness-of-fit measures

| Name | Definition |
|---|---|
| $\rho^2$ | Log-Likelihood ratio<br>The log-likelihood ratio is one minus the ratio of the full-model log-likelihood to the intercept-only log-likelihood<br>$$1 - \frac{LL_{Full}}{LL_{Null}}$$ |
| Count $R^2$ | The predicted values that match actual values constitute the total number of correct predictions. The Count $R^2$ is this correct count divided by the total count.<br>$$\frac{Total\ Correctly\ Predicted\ Count}{Total\ Count}$$ |
| Precision | Precision is the fraction of the relevant instances among the retrieved instances.<br>$$\frac{True\ Positives}{True\ Positives + False\ Positives}$$ |
| Recall | The recall is the fraction of the relevant instances that were retrieved.<br>$$\frac{True\ Positives}{True\ Positives + False\ Negatives}$$ |
| $F_1$ Score | $F_1$ Score is a measure of the test accuracy and combines precision and recall and is described as the harmonic mean of precision and recall<br>$$2 * \frac{Precision * Recall}{Precision + Recall}$$ |
| Accuracy | Accuracy is the total number of correct predictions overall predictions<br>$$\frac{Number\ of\ Correct\ Predictions}{Size\ of\ the\ Dataset}$$ |

# Appendix B

For the estimation, we consider "Regular Car" as the base alternative and present the estimated coefficients for "Private AVs" (PAV) and "Shared AVs" (SAV).

**Table 12**
Estimated coefficients for socio-demographic characteristics.

| Variables | Levels | Proposed | | Ver_LK | | Ver_LKOE | | Ver_OE | |
|---|---|---|---|---|---|---|---|---|---|
| | | PAV | SAV | PAV | SAV | PAV | SAV | PAV | SAV |
| | Constant | 0.183 | -0.237 | -0.829 | -0.488 | 0.456 | -0.522 | 0.495 | -0.175 |
| | Male | 0.035 | 0.042 | 0.013 | -0.293 | 0.037 | 0.031 | 0.084 | 0.023 |
| Age | 16 and 25 | 0.566 | 0.327 | 0.23 | -0.326 | 0.441 | 0.51 | 0.556 | 0.304 |
| | 26 and 35 | 0.463 | 0.179 | 0.547 | -0.189 | 0.279 | 0.572 | 0.499 | 0.039 |
| | 36 and 45 | 0.343 | 0.059 | 0.223 | -0.215 | 0.379 | 0.543 | 0.313 | -0.111 |
| | 46 and 55 | 0.292 | 0.198 | 0.242 | -0.114 | 0.559 | 0.983 | 0.201 | -0.03 |
| Household Income | 0 - $24,999 | -0.047 | 0.228 | -0.045 | -0.102 | -0.041 | 0.362 | -0.065 | 0.524 |
| | $25,000 and $49,999 | -0.013 | 0.098 | -0.129 | -0.295 | 0.261 | 0.699 | -0.152 | -0.161 |
| | $50,000 and $74,999 | 0.084 | 0.147 | -0.032 | -0.462 | 0.477 | 0.836 | -0.158 | -0.053 |
| | $75,000 and $99,999 | -0.082 | 0.223 | 0.093 | 0.131 | 0.468 | 0.944 | -0.69 | -0.332 |
| Educational Qualification | Less than high school graduate | -0.806 | -0.865 | -0.558 | -0.545 | 0.325 | -0.105 | -1.126 | -0.992 |
| | High school graduate or GED | -0.359 | -0.513 | -0.234 | -0.384 | 0.115 | 0.091 | -0.351 | -0.721 |
| | Some college or associate degree | -0.298 | -0.48 | 0.276 | -0.436 | -0.346 | -0.508 | -0.369 | -0.243 |
| | Bachelors' degree | -0.151 | -0.033 | 0.145 | 0.028 | 0.093 | 0.319 | -0.365 | -0.104 |
| Ethnicity | European American | -0.129 | 0.01 | 0.273 | 0.169 | -0.325 | -0.275 | -0.286 | 0.076 |
| | Black or African American | -0.096 | -0.267 | 0.41 | 0.031 | -0.472 | -0.474 | -0.242 | -0.327 |
| | American Indian or Alaska Native | 0.264 | -0.801 | 0.181 | -1.159 | -0.148 | -0.232 | 0.398 | -0.906 |
| | Asian | -0.154 | -0.026 | -0.329 | 0.297 | -0.339 | -1.135 | -0.091 | 0.033 |
| | Native Hawaiian or other Pacific Islander | -0.293 | -0.755 | 0.195 | 0.385 | 0.004 | -0.523 | -0.6 | -1.306 |
| Employment | Full-time | -0.285 | -0.215 | -0.294 | 0.165 | -0.181 | -0.46 | -0.088 | -0.088 |
| | Part-time | -0.358 | -0.292 | -0.043 | 0.199 | -0.352 | -0.21 | -0.457 | -0.677 |
| Number of adults | | -0.122 | -0.024 | 0.049 | 0.12 | -0.198 | -0.148 | -0.17 | 0.014 |
| Number of children aged between 8 and 17 | | 0.022 | 0.092 | 0.055 | 0.099 | -0.153 | -0.097 | 0.101 | 0.178 |
| Number of children aged less than 8 | | -0.11 | -0.184 | -0.165 | -0.254 | 0.079 | -0.005 | -0.158 | -0.28 |

Coefficients at 99% confidence level  Coefficients at 95% confidence level  Coefficients at 90% confidence level

**Table 13**
Estimated coefficients for travel, familiarity with AVs and SP.

| Variables | Levels | Proposed | | Ver_LK | | Ver_LKOE | | Ver_OE | |
|---|---|---|---|---|---|---|---|---|---|
| | | PAV | SAV | PAV | SAV | PAV | SAV | PAV | SAV |
| Total miles travelled is less than 35 miles | | 0.016 | -0.2 | 0.183 | -0.405 | -0.038 | -0.359 | -0.034 | 0.064 |

| | | | | | | | | |
|---|---|---|---|---|---|---|---|---|
| Total miles travelled is between 35 and 70 miles | | -0.102 | -0.22 | 0.014 | -0.268 | -0.058 | -0.143 | -0.201 | -0.167 |
| Mode used for commute | Walk | -0.04 | 0.161 | 0.02 | 0.149 | 0.001 | 0.544 | -0.023 | 0.015 |
| | Bike | 0.036 | 0.697 | -0.334 | 0.775 | -0.024 | 0.191 | 0.09 | 0.876 |
| | Motorcycle/moped | 0.101 | -0.077 | -0.794 | -1.458 | 0.133 | 0.015 | 0.223 | -0.698 |
| | Car/SUV/Van/Pickup | -0.271 | -0.117 | -0.264 | -0.602 | -0.416 | -0.288 | 0.147 | 0.129 |
| | Car-sharing | 0.181 | -0.318 | 0.217 | -0.036 | -0.331 | -0.778 | 0.494 | -0.045 |
| | Taxi/ride-hailing | 0.145 | 0.185 | 0.391 | 0.07 | -0.296 | -0.035 | 0.106 | 0.447 |
| | Public transport | 0.189 | 0.27 | 0.053 | 0.454 | -0.251 | -0.088 | 0.37 | 0.47 |
| Does not own a car | | -0.275 | 0.039 | -0.697 | -0.283 | -0.159 | 0.453 | -0.114 | -0.449 |
| Travels alone in the car | | 0.132 | -0.214 | -0.417 | -0.572 | 0.069 | -0.063 | 0.238 | -0.112 |
| Travels with others in the car | | 0.396 | 0.24 | 0.003 | 0.089 | 0.117 | 0.44 | 0.311 | 0.019 |
| Leaving items in the car is important | | 0.044 | 0.274 | 0.093 | 0.229 | 0.419 | 0.486 | -0.041 | 0.262 |
| Familiarity with AV | | 0.021 | 0.218 | 0.087 | 0.602 | 0.057 | -0.134 | -0.144 | 0.239 |
| Has ridden AV | | 0.347 | -0.081 | 0.216 | -0.702 | 0.642 | 0.412 | 0.204 | -0.059 |
| Purchase Cost (Regular Car) | | 0.848 | 0.17 | 0.751 | -0.087 | 1.022 | 0.374 | 0.611 | 0.023 |
| Purchase Cost (Private AV) | | -1.126 | -0.34 | -0.93 | -0.132 | -1.264 | -0.37 | -0.955 | -0.259 |
| Membership Cost (Shared AV) | | 0.067 | -0.26 | 0.08 | -0.218 | 0.096 | -0.293 | 0.072 | -0.265 |
| Travel Cost (Regular Car) | | 0.372 | 0.254 | 0.445 | 0.411 | 0.326 | 0.334 | 0.333 | -0.067 |
| Travel Cost (Private AV) | | -0.361 | -0.106 | -0.206 | -0.18 | -0.428 | -0.12 | -0.403 | 0.06 |
| Travel Cost (Shared AV) | | 0.012 | -0.447 | -0.035 | -0.385 | 0.071 | -0.461 | 0 | -0.43 |
| Parking Charge (Regular Car) | | 0.205 | 0.046 | 0.227 | -0.12 | 0.189 | -0.025 | 0.203 | 0.097 |
| Parking Charge (Private AV) | | -0.227 | -0.066 | -0.165 | -0.021 | -0.255 | -0.062 | -0.234 | -0.016 |
| Attitudinal_1 | | 0.045 | -1.065 | -0.102 | 1.32 | 0.402 | -0.871 | 0.785 | -0.385 |
| Attitudinal_2 | | 2.124 | 1.306 | -2.103 | -1.1 | -2.382 | -2.002 | 0.827 | 1.11 |
| Coefficients at 99% confidence level | | | Coefficients at 95% confidence level | | | | Coefficients at 90% confidence level | | |

In this research, attitudes were estimated as a latent variable (Coe_1 and Coe_2), which was later included as an explanatory variable in the utility equation for mode choice. Hence, while assessing the effects of attitudes in Table 14 and Table 15, we consider the signs and magnitudes of variables "Attitudinal_1" and "Attitudinal_2" from Table 13.

**Table 14**
Estimated coefficients for Likert scale responses

| Variable | Level | Ver_LK | | | | Ver_LKOE | | | |
|---|---|---|---|---|---|---|---|---|---|
| | | Prop | | Ind | | Prop | | Ind | |
| | | Coe_1 | Coe_2 | Coe_1 | Coe_2 | Coe_1 | Coe_2 | Coe_1 | Coe_2 |
| Constant_1 | | 0.423 | -0.389 | 0.045 | 0.542 | 0.682 | -0.771 | 0.212 | -0.232 |
| Learning to use Autonomous Vehicles will be easy for me (PEoU_1) | PEoU_11 | 0.259 | -0.026 | -0.31 | 0.499 | -0.022 | 0.218 | 0.126 | 0.498 |
| | PEoU_12 | 0.483 | 0.115 | -0.306 | 0.052 | 0.191 | 0.134 | 0.189 | 0.068 |
| | PEoU_13 | -0.027 | 0.218 | 0.118 | -0.053 | 0.044 | 0.087 | 0.026 | 0.05 |
| | PEoU_14 | 0.105 | 0.311 | 0.073 | -0.165 | -0.037 | 0.312 | -0.031 | -0.127 |
| | PEoU_15 | 0.038 | 0.245 | 0.104 | -0.05 | -0.281 | 0.39 | -0.115 | -0.35 |
| I will find it easy to get Autonomous Vehicles to do what I want them to do (PEoU_2) | PEoU_21 | 0.072 | -0.013 | -0.497 | 0.059 | -0.118 | 0.05 | 0.046 | 0.25 |
| | PEoU_22 | -0.192 | -0.276 | 0.148 | 0.049 | -0.434 | -0.212 | -0.372 | 0.264 |
| | PEoU_23 | -0.087 | -0.286 | 0.146 | 0.11 | -0.073 | -0.151 | 0.004 | 0.293 |
| | PEoU_24 | -0.319 | -0.034 | 0.382 | -0.167 | -0.294 | -0.021 | -0.135 | 0.078 |

| | | | | | | | | |
|---|---|---|---|---|---|---|---|---|
| | PEoU_25 | -0.144 | -0.668 | 0.246 | 0.502 | 0.233 | -0.148 | 0.259 | 0.35 |
| It will be easy for me to become skilful at using Autonomous Vehicles (PEoU_3) | PEoU_31 | -0.51 | -0.365 | 0.171 | 0.212 | -0.291 | 0.252 | -0.42 | -0.024 |
| | PEoU_32 | -0.798 | -0.551 | 0.614 | 0.015 | 0.287 | -0.389 | 0.096 | 0.179 |
| | PEoU_33 | -0.038 | -0.413 | -0.14 | -0.162 | 0.39 | -0.282 | 0.166 | 0.034 |
| | PEoU_34 | -0.059 | -0.414 | -0.139 | -0.122 | 0.378 | -0.341 | 0.169 | 0.128 |
| | PEoU_35 | 0.205 | -0.128 | -0.443 | -0.432 | -0.114 | -0.371 | -0.135 | 0.003 |
| I will find Autonomous Vehicles easy for me (PEoU_4) | PEoU_41 | -0.223 | 0.544 | 0.67 | 0.176 | 0.013 | -0.084 | -0.141 | 0.377 |
| | PEoU_42 | 0.789 | 0.044 | -0.531 | 0.098 | -0.164 | -0.06 | -0.001 | -0.082 |
| | PEoU_43 | 0.24 | -0.109 | -0.079 | 0.228 | -0.268 | -0.083 | -0.186 | -0.041 |
| | PEoU_44 | 0.089 | 0.115 | 0.079 | 0.017 | 0.037 | -0.243 | 0.027 | 0.184 |
| | PEoU_45 | 0.511 | 0.413 | -0.27 | -0.418 | 0.95 | -0.02 | 0.772 | 0.296 |
| Using Autonomous Vehicles will be useful in meeting my travel needs (PU_1) | PU_11 | -0.023 | -0.54 | 0.111 | 1.031 | 0.182 | -0.248 | 0.072 | 1.298 |
| | PU_12 | 0.433 | -0.133 | -0.304 | 0.457 | 0.361 | -0.099 | 0.311 | 1.027 |
| | PU_13 | 0.182 | 0.295 | -0.075 | -0.114 | -0.294 | 0.67 | -0.178 | -0.085 |
| | PU_14 | 0.311 | 0.757 | -0.161 | -0.62 | -0.384 | 1.27 | -0.063 | -0.66 |
| | PU_15 | 0.296 | 0.804 | -0.161 | -0.65 | -0.713 | 1.227 | -0.325 | -0.701 |
| Autonomous Vehicles will let us do other tasks such as eating, watching a movie, be on a cell phone during my trip (PU_2) | PU_21 | -0.4 | 0.085 | 0.307 | -0.188 | 0.052 | 0.265 | 0.326 | 0.009 |
| | PU_22 | -0.23 | -0.307 | 0.096 | 0.168 | -0.374 | 0.249 | -0.231 | -0.314 |
| | PU_23 | 0.217 | -0.214 | -0.405 | 0.009 | -0.186 | 0.136 | 0.021 | -0.059 |
| | PU_24 | 0.078 | -0.158 | -0.231 | -0.014 | -0.246 | 0.148 | -0.04 | -0.129 |
| | PU_25 | -0.042 | -0.293 | -0.035 | 0.084 | -0.338 | 0.191 | -0.146 | -0.131 |
| Using Autonomous Vehicles will decrease my accident risk (PU_3) | PU_31 | 0.432 | 0.142 | -0.717 | 0.274 | -0.047 | -0.174 | -0.098 | 0.685 |
| | PU_32 | -0.196 | 0.345 | 0.033 | 0.032 | -0.203 | 0.085 | -0.28 | 0.486 |
| | PU_33 | 0.079 | 0.479 | -0.197 | -0.132 | 0.059 | 0.363 | 0.052 | 0.248 |
| | PU_34 | -0.313 | 0.623 | 0.265 | -0.302 | -0.326 | 0.891 | -0.174 | -0.451 |
| | PU_35 | -0.287 | 0.751 | 0.233 | -0.386 | -0.084 | 0.755 | 0.153 | -0.216 |
| Using Autonomous Vehicles will relieve my stress of driving (PU_4) | PU_41 | 0.255 | -0.213 | -0.347 | 0.29 | 0.163 | -0.311 | -0.041 | 0.563 |
| | PU_42 | -0.232 | -0.363 | 0.007 | 0.528 | 0.103 | -0.166 | -0.038 | 0.459 |
| | PU_43 | -0.216 | 0.063 | 0.081 | 0.059 | -0.086 | 0.155 | 0.048 | 0.037 |
| | PU_44 | -0.515 | 0.386 | 0.337 | -0.254 | -0.08 | 0.007 | -0.022 | 0.177 |
| | PU_45 | -0.706 | 0.266 | 0.658 | -0.235 | -0.136 | 0.429 | 0.103 | -0.447 |
| I find Autonomous Vehicles to be useful when I'm impaired (PU_5) | PU_51 | -0.015 | -0.056 | -0.185 | -0.022 | -0.085 | -0.124 | 0.094 | 0.388 |
| | PU_52 | 0.173 | -0.081 | -0.451 | -0.099 | 0.106 | -0.047 | 0.239 | 0.021 |
| | PU_53 | -0.112 | -0.212 | -0.015 | 0.104 | -0.091 | 0.089 | 0.152 | -0.119 |
| | PU_54 | -0.321 | 0.168 | 0.118 | -0.358 | -0.209 | 0.155 | 0.031 | -0.302 |
| | PU_55 | -0.673 | 0.352 | 0.483 | -0.512 | -0.6 | 0.24 | -0.318 | -0.455 |

Coefficients at 99% confidence level  Coefficients at 95% confidence level  Coefficients at 90% confidence level

**Table 15**
Estimated coefficients for the topics

| Variable | Level | Ind | | Prop | |
|---|---|---|---|---|---|
| | | Coe_1 | Coe_2 | Coe_1 | Coe_2 |
| Constant_1 | | 0.005 | 0.145 | 0.048 | -0.052 |
| Do you think that it will be easy to use Autonomous Vehicles? Explain why. | To_L11 | 0.038 | 0.055 | 0.086 | 0.063 |
| | To_L12 | -0.071 | -0.083 | -0.198 | -0.087 |
| | To_L13 | -0.02 | 0.127 | 0.12 | 0.28 |
| | To_L14 | -0.018 | -0.059 | -0.049 | -0.06 |
| Do you believe that Autonomous Vehicles are useful? Explain why. | To_L21 | -0.07 | 0.047 | -0.019 | 0.148 |
| | To_L22 | 0.052 | 0.263 | 0.351 | 0.318 |
| | To_L23 | -0.075 | 0.155 | 0.081 | 0.321 |
| | To_L24 | -0.005 | -0.046 | -0.059 | -0.059 |
| | To_L25 | -0.133 | 0.027 | -0.103 | 0.15 |
| | To_L26 | 0.036 | -0.18 | -0.155 | -0.342 |
| | To_L27 | -0.039 | 0.028 | -0.002 | 0.059 |

Coefficients at 99% confidence level    Coefficients at 95% confidence level    Coefficients at 90% confidence level

# Appendix C (Python Code for Topic Models)

**Topic Model Analysis**

The code is written to extract topics from text data. The code supports performing the following analysis:-

1. Latent Dirichlet Allocation (LDA)
2. Supervised Latent Dirichlet Allocation (sLDA)

Latent Dirichlet allocation can be performed using Gensim or Tomotopy. Supervised LDA can be performed using Tomotopy. The dependent variable can be linear or binary. Visualisations to evaluate the results from the Topic Models, pyLDAvis to understand topics and the inter-topic distance.

**Importing the Libraries**

```
### ************************ Importing Packages ************************ ###
from __future__ import division
import re                    # regular expressions
import numpy as np           # scientific computing
import pandas as pd          # datastructures and computing
import pprint as pprint      # better printing
import os
import os.path
```

**# Gensim**

```
import gensim
import gensim.corpora as corpora
from gensim.utils import simple_preprocess
from gensim.models import CoherenceModel
```

**# Lemmatization**

```
from nltk.stem import PorterStemmer
from nltk.tokenize import sent_tokenize, word_tokenize
```

**# Plotting tools**

```
import pyLDAvis              # interactive Topic Model visualisation
import pyLDAvis.gensim
import matplotlib.pyplot as plt
```

**# Libraries for Topic Models**

```
import sys
import tomotopy as tp
```

**# fix random generator seed (for reproducibility of results)**

```
np.random.seed(42)
import warnings
warnings.filterwarnings("ignore", category=DeprecationWarning)
```

**Creating the list of Stop Words**

**# NLTK Stop words**

```
from nltk.corpus import stopwords
stop_words = stopwords.words('english')
stop_words.extend(['also', 'back', 'cant', 'come', 'could', 'done', 'dont', 'due', 'els', 'etc', 'hope', 'howev','know',
          'let', 'like', 'may', 'mayb', 'much', 'must', 'new', 'non', 'one', 'other', 'plu', 'pretti', 'said',
          'say', 'see', 'sinc', 'someon', 'someoth', 'therefor', 'today', 'want', 'well', 'would', 'ye', 'car',
          'cars', 'think', 'autonomous', 'vehicle', 'vehicles','people', 'seem', 'seems', 'really', 'still',
          'however', 'believe', 'right', 'truly', 'automatic', 'sound', 'sounds', 'general', 'become', 'total',
          'totally', 'tell', 'something', 'anything', 'person', 'phone'])
```

**Importing the Dataset**

The dataset preparation is quite important here. If the model used is simple LDA, only the text data is mandatory. However, for the supervised LDA, it requires the response variable (dependent variable) along with the text data.

```
### ************************ Importing Datasets ************************ ###
```

```
directo = "<folder_path>"
df = pd.read_excel(directo + "\\<filename>")
df.head()
|===========================================================================
```
# convert the content field in dataset into a list
```
data = df.AVO_EUsT.values.tolist()
resp = df.Resp.values.tolist()
data[:5]
resp[:5]
```

**Cleaning the Dataset**

In this section of the code, the following data cleaning techniques are used:-

1. E-mail id and Newline characters
2. Remove "StopWords" from the dataset
3. Forming bigrams and trigrams
4. Stemming

```
### ************************ Datasets Cleaning ************************ ###
```
**E-mail id and New line characters**

# Remove Emails
```
data = [re.sub('\S*@\S*\s?', '', sent) for sent in data]
```
# Remove new line characters
```
data = [re.sub('\n', ' ', sent) for sent in data]
pprint.pprint(data[:5])
```

**Replacing Phrases with Meaningful Words**
```
data_1 = []
p1  = re.compile("(do\s*not\s*trust|don't\s*trust|don't\s*fully\s*trust|would\s*\not*\s*trust|never\s*trust)")
p2  = re.compile("unsafe|not\s*safe|not\s*feel\s*safe|not\s*be\s*safe|doesn't\s*seem\s*very\s*safe|\
don't\s*feel\s*it's\s*safe")
p3  = re.compile("not\*feel")
p4  = re.compile("don't\*think")
for item in data:
    data_1a = p1.sub("no_trust", item)
    data_1b = p2.sub("unsafe", data_1a)
    data_1c = p3.sub("not_feel", data_1b)
    data_1d = p2.sub("dont_think", data_1c)
    data_1.append(data_1d)
data =  data_1
print(data[:5])
```
**Removing StopWords**
```
Function to remove StopWords
def remove_stopwords(texts):
    """
    objective:
       function to remove stopwords from the paragraph/sentence
       uses the preprocess
    input:
       paragraph/sentences
    output:
       wordlist after the stopwords are removed
    """
    return [[word for word in simple_preprocess(str(doc)) if word not in stop_words]
            for doc in texts]
data_words_nostops = remove_stopwords(data)
data_words_nostops[:5]
```
**Forming Bigrams and Trigrams**

```python
def make_bigrams(texts):
    """
    objective:
        takes the processed text- after preprocessing and stop word removal
    input:
        preprocessed text
    output:
        text with bigrams
    """
    return [bigram_mod[text] for text in texts]
def make_trigrams(texts):
    """
    objective:
        generate trigrams for the text
    input:
        text with bigrams
    output:
        text with trigrams
    """
    return [trigram_mod[bigram_mod[text]] for text in texts]
```

# Build functions to remove stopwords, bigram and trigram models- calibration dataset

```python
bigram = gensim.models.phrases.Phrases(data, min_count=5, threshold=100)
trigram = gensim.models.phrases.Phrases(bigram[data], threshold=100)
# Passing the parameters to the bigram/trigram- calibration dataset

bigram_mod = gensim.models.phrases.Phraser(bigram)
trigram_mod = gensim.models.phrases.Phraser(trigram)
data_words_bigrams = make_bigrams(data_words_nostops)
data_words_bigrams[:5]
```

**Lemmatization**

```python
ps = PorterStemmer()
data_lemmatized = []
for texts in data_words_bigrams:
    data_lemmatized.append([ps.stem(doc) for doc in texts])
data_lemmatized[:5]
```

**Writing the Files to the dataset**

```python
df['Cleaned_Data'] = data_lemmatized
df.head()
df.to_csv(directo + "\\Output_Q1_Words.csv")
```

**LDA Model**

# Defining the LDA Function

```python
def lda_model(input_list, save_path):
    """
    desc:
        the function estimates the LDA model and outputs the estimated topics
    input:
        list with documents as responses
    output:
        prints the topics
        words and their corresponding proportions
    """
    mdl = tp.LDAModel(tw=tp.TermWeight.ONE,      # Term weighting
                min_cf=3,              # Minimum frequency of words
                rm_top=0,              # Number of top frequency words to be removed
                k=4,                   # Number of topics
                seed=42)
    for n, line in enumerate(input_list):
        ch = " ".join(line)
```

```python
        docu = ch.strip().split()
        mdl.add_doc(docu)
    mdl.burn_in = 10000
    mdl.train(10000)
    print('Num docs: ', len(mdl.docs), 'Vocab size: ', mdl.num_vocabs, 'Num words: ', mdl.num_words)
    print('Removed words: ', mdl.removed_top_words)
    print('Training...', file=sys.stderr, flush=True)
    for i in range(0, 50000, 10):
        mdl.train(100)
        print('Iteration: {}\tLog-likelihood: {}'.format(i, mdl.ll_per_word))
    print('Saving...', file=sys.stderr, flush=True)
    mdl.save(save_path, True)
    for k in range(mdl.k):
        print('Topic #{}'.format(k))
        for word, prob in mdl.get_topic_words(k):
            print('\t', word, prob, sep='\t')
    return mdl
```

**Estimating the Topic Model**

```python
print('Running LDA')
lda_model = lda_model(data_lemmatized, 'test.lda_4_T.bin')
```

**Supervised LDA**

```python
def slda_model(documents, dep_var, save_path):
    """
    desc:
        the function estimates the sLDA model and outputs the estimated topics
    input:
        list with documents as responses
        dependent variable
    output:
        prints the topics
        words and their corresponding proportions
    """
    smdl = tp.SLDAModel(tw=tp.TermWeight.ONE,        # Term weighting
                min_cf=3,              # Minimum frequency of words
                rm_top=0,              # Number of top frequency words to be removed
                k=4,                   # Number of topics
                vars=['b'],            # Number of dependent variables
                seed=42)
    for row, pred in zip(documents, dep_var):
        pred_1 = []
        pred_1.append(pred)
        ch = " ".join(row)
        docu = ch.strip().split()
        smdl.add_doc(words=docu, y=pred_1)
    smdl.burn_in = 10000
    smdl.train(10000)
    # Printing the output statistics
    print('Num docs: ', len(smdl.docs), 'Vocab size: ', smdl.num_vocabs, 'Num words: ', smdl.num_words)
    print('Removed top words: ', smdl.removed_top_words)
    print('Training...', file=sys.stderr, flush=True)
    for i in range(0, 50000, 10):
        smdl.train(100)
        print('Iteration: {}\tLog-likelihood: {}'.format(i, smdl.ll_per_word))
    print('Saving...', file=sys.stderr, flush=True)
    smdl.save(save_path, True)
    for k in range(smdl.k):
        print('Topic #{}'.format(k))
        for word, prob in smdl.get_topic_words(k):
```

```
        print('\t', word, prob, sep='\t')
    return smdl
print('Running Supervised LDA')
slda_model = slda_model(data_lemmatized, resp, 'test.slda_4_T.bin')
```

**Visualising the Results of LDA**

pyLDAvis does not have a module that allows Topic Models estimated using Tomotopy to be used directly for plotting the graphs. It however allows plotting after the following parameters are computed for each of the Topic Models:-

1. Phi
    a. probabilities of each word(W) for a given topic(K) under consideration
    b. is a K x W vector
2. theta
    a. probability mass function over ``K'' topics for all the documents in the corpus (D)
    b. is a D x K matrix
3. n(d)
    a. number of tokens for each document
4. vocab
    a. vector of terms in the vocabulary
    b. presented in the same order as in ``phi''
5. M(w)
    a. frequency of term ``w'' across the entire corpus

**Computing the value of ``Phi'' for the Model**

```
def compute_phi(model):
    """
    desc:
        this function computes the value of phi for visualising the results of Topic Model
        probabilities of each word for a given topic
    input:
        the Topic Model
    output:
        K x W vector
        K = number of topics
        W = number of words
    """
    mat_phi1 = []
    for i in range(model.k):
        mat_phi1.append(model.get_topic_words(i,model.num_vocabs))
    list_words = []
    for text in mat_phi1[0]:
        list_words.append(text[0])
    list_words.sort()
    mat_phi2 = [[i * j for j in range(model.num_vocabs)] for i in range(model.k+1)]
    for i in range(model.num_vocabs):
        mat_phi2[0][i] = list_words[i]
    j1 = []
    k1 = []
    m = 0
    while m < model.k:
        j1.append(m)
        m += 1
    n = 1
    while n <= model.k:
```

```
            k1.append(n)
            n += 1
        for j, k in zip(j1, k1):
            for index, word in enumerate(mat_phi2[0]):
                #print(word)
                for item in mat_phi1[j]:
                    #print(item)
                    if word == item[0]:
                        mat_phi2[k][index] = item[1]
        if os.path.isfile(directo + '\\topic_word_prob_lda_4_T.csv'):
            with open(directo + '\\topic_word_prob_slda_4_T.csv', 'w') as f:
                for item in mat_phi2:
                    for items in item:
                        f.writelines("%s, " % items)
                    f.writelines("\n")
                f.close()
        else:
            with open(directo + '\\topic_word_prob_lda_4_T.csv', 'w') as f:
                for item in mat_phi2:
                    for items in item:
                        f.writelines("%s, " % items)
                    f.writelines("\n")
                f.close()
        return mat_phi2[0], mat_phi2[1:]
```

**Computing the value of ``Theta'' for the Model**

**For LDA Model**

```
def compute_theta_lda(model, data):
    """
    desc:
        this function computes the value of theta for visualising the results of Topic Model
        probabilities mass function over "K" topics for all documents (D) in the corpus
    input:
        the Topic Model
        dataset
    output:
        D x K vector
        D = number of documents
        K = number of topics
    """
    mat_theta = []
    for n, line in enumerate(data):
        ch = " ".join(line)
        docu = ch.strip().split()
        theta_val = model.infer(doc=model.make_doc(docu),
                        iter=100,
                        workers=0,
                        together=False)
        mat_theta.append(theta_val[0])
    with open(directo + '\\topic_probabilities_lda_4_T.csv', 'w') as f:
        for item in mat_theta:
            for items in item:
                f.writelines("%s, " %items)
            f.writelines("\n")
        f.close()
    return mat_theta
```

**For sLDA Model**

```
def compute_theta_slda(model, data, dep_var):
    """
    desc:
```

```
        this function computes the value of theta for visualising the results of Topic Model
        probabilities mass function over "K" topics for all documents (D) in the corpus
    input:
        the Topic Model
        dataset
        dependent variable
    output:
        D x K vector
        D = number of documents
        K = number of topics
    """
    mat_theta = []
    for line, dep in zip(data, dep_var):
        pred_1 = []
        pred_1.append(dep)
        ch = " ".join(line)
        docu = ch.strip().split()
        theta_val = model.infer(doc=model.make_doc(words=docu, y=pred_1),
                            iter=100,
                            workers=0,
                            together=False)
        mat_theta.append(theta_val[0])
    with open(directo + '\\topic_probabilities_slda_4_T.csv', 'w') as f:
        for item in mat_theta:
            for items in item:
                f.writelines("%s, " %items)
            f.writelines("\n")
        f.close()
    return mat_theta
```

**Number of Tokens per document**

```
def num_token(data):
    """
    desc:
        this function computes number of tokens per document for the entire corpus
    input:
        dataset
    output:
        N x 1 vector
        N = number of tokens in the document
    """
    numb_tok = []
    for text in data:
        numb_tok.append(len(text))
    return numb_tok
```

**Frequency of Words in the Corpus**

```
def freq_words(vocabs, data):
    """
    desc:
        this function computes the frequency of words in the entire corpus
    input:
        list of words
        dataset
    output:
        N x 1 vector
        N = frequency of words in the document
    """
    fre_words = []
    for words in vocabs:
        words_freq = 0
```

```
        for line in data:
            for ind_words in line:
                if words == ind_words:
                    words_freq += 1
        fre_words.append(words_freq)
    return fre_words
```

**Visualising the Results of LDA Model**

**Computing the Parameters for Visualising LDA Model**

**# Loading the LDA model**

```
lda_model = tp.LDAModel.load('test.lda_4_T.bin')
#lda_model.get_topic_word_dist(2)
lvocab, lphi_val = compute_phi(lda_model)
ltheta_val = compute_theta_lda(lda_model, data_lemmatized)
lnum_token = num_token(data_lemmatized)
lfreq_terms = freq_words(lvocab, data_lemmatized)
```

**Plotting in pyLDAvis (LDA)**

**# Visualising the Results**

```
pyLDAvis.enable_notebook()
data_lda = {'topic_term_dists': lphi_val,
        'doc_topic_dists' : ltheta_val,
        'doc_lengths'     : lnum_token,
        'vocab'           : lvocab,
        'term_frequency'  : lfreq_terms}
print('Topic-Term shape: %s' % str(np.array(data_lda['topic_term_dists']).shape))
print('Doc-Topic shape: %s' % str(np.array(data_lda['doc_topic_dists']).shape))
vis_lda = pyLDAvis.prepare(**data_lda)
pyLDAvis.display(vis_lda)
```

**Visualising the Results of Supervised LDA Model**

**Computing the Parameters for Visualising Supervised LDA Model**

**# Loading the sLDA model**

```
slda_model = tp.SLDAModel.load('test.slda_4_T.bin')
svocab, sphi_val = compute_phi(slda_model)
stheta_val = compute_theta_slda(slda_model, data_lemmatized, resp)
snum_token = num_token(data_lemmatized)
sfreq_terms = freq_words(svocab, data_lemmatized)
```

**Plotting in pyLDAvis (sLDA)**

**# Visualising the Results**

```
pyLDAvis.enable_notebook()
data_slda = {'topic_term_dists': sphi_val,
        'doc_topic_dists' : stheta_val,
        'doc_lengths'     : snum_token,
        'vocab'           : svocab,
        'term_frequency'  : sfreq_terms}
print('Topic-Term shape: %s' % str(np.array(data_slda['topic_term_dists']).shape))
print('Doc-Topic shape: %s' % str(np.array(data_slda['doc_topic_dists']).shape))
vis_slda = pyLDAvis.prepare(**data_slda)
pyLDAvis.display(vis_slda)
```

**Computing Scores for use in Estimation**

In this portion of the code, values are computed for each document in the corpus. The values are computed based on the words used in each of the documents in the corpus. Scores will be computed for each topic. This will be based on the probability values in each of the topics.

```
def compute_scores(list_dataset, list_word_prob):
    """
    desc:
        this function will take the cleaned dataset and list of word probabilities per topic and compute the scores
```

```
input:
    cleaned dataset as a list
    word probabilities as a dataframe
output:
    scores for each document in the corpus
"""
n = len(list_dataset)
prob_list = [[0 for i in range(5)] for i in range(n)]
for index, document in enumerate(list_dataset):
    # remember to change the number of variables based on the number of topics
    probab_1 = 0
    probab_2 = 0
    probab_3 = 0
    probab_4 = 0
    for word in document:
        for index1, row in list_word_prob.iterrows():
            item  = row['Word']
            prob1 = row['Prob_1']
            prob2 = row['Prob_2']
            prob3 = row['Prob_3']
            prob4 = row['Prob_4']
            if word == item:
                probab_1 += prob1
                probab_2 += prob2
                probab_3 += prob3
                probab_4 += prob4
        prob_list[index][0] = probab_1
        prob_list[index][1] = probab_2
        prob_list[index][2] = probab_3
        prob_list[index][3] = probab_4
        prob_list[index][4] = probab_1 + probab_2 + probab_3 + probab_4
return prob_list
```

**Computing the Scores for LDA**

```
lda_dist = pd.read_csv(directo + "\\topic_word_prob_lda_4_T.csv", header=None)
lda_distT = lda_dist.T
lda_distT.columns = ['Word', 'Prob_1', 'Prob_2', 'Prob_3', 'Prob_4']
lda_distT['Word']   = lda_distT['Word'].str.strip()
lda_distT['Prob_1'] = pd.to_numeric(lda_distT.Prob_1, errors='coerce')
lda_distT['Prob_2'] = pd.to_numeric(lda_distT.Prob_2, errors='coerce')
lda_distT['Prob_3'] = pd.to_numeric(lda_distT.Prob_3, errors='coerce')
lda_distT['Prob_4'] = pd.to_numeric(lda_distT.Prob_4, errors='coerce')
probab_lda = compute_scores(data_lemmatized, lda_distT)
df['probab_lda'] = probab_lda
```

**Computing the Scores for sLDA**

```
slda_dist = pd.read_csv(directo + "\\topic_word_prob_slda_4_T.csv", header=None)
slda_distT = slda_dist.T
slda_distT.columns = ['Word', 'Prob_1', 'Prob_2', 'Prob_3', 'Prob_4']
slda_distT['Word']   = slda_distT['Word'].str.strip()
slda_distT['Prob_1'] = pd.to_numeric(slda_distT.Prob_1, errors='coerce')
slda_distT['Prob_2'] = pd.to_numeric(slda_distT.Prob_2, errors='coerce')
slda_distT['Prob_3'] = pd.to_numeric(slda_distT.Prob_3, errors='coerce')
slda_distT['Prob_4'] = pd.to_numeric(slda_distT.Prob_4, errors='coerce')
probab_slda = compute_scores(data_lemmatized, slda_distT)
df['probab_slda'] = probab_slda

df.to_csv(directo + "\\Open_Ended_Q1_Scores_4_Topic.csv")
```

# Appendix D (Pyro Code for the Framework to Measure Attitudes)

**Contents**

1. Problem Description
2. Data Preparation
3. Probabilistic Graphical Model and the Generative Process
4. Proposed Model

**Problem Description**

We use the data on the mode choice for commute trips by students and workers from the USA. A stated-preference (SP) survey was used. The questionnaire collected information on:-

1. Socio-demographic characteristics
2. Travel characteristics
3. Familiarity with Autonomous Vehicles
4. Attitudes
5. SP attributes

The attitudes were measured using 5-point Likert scales. For some attitudes, open-ended questions were also presented to the respondents.

**Importing the Libraries**

```
import numpy as np
import pandas as pd
from matplotlib import pyplot as plt
import seaborn as sns
import torch
torch.set_default_tensor_type("torch.cuda.FloatTensor")
import pyro
import pyro.distributions as dist
from pyro.contrib.autoguide import AutoDiagonalNormal, AutoMultivariateNormal
from pyro.infer import MCMC, NUTS, HMC, SVI, Trace_ELBO
from pyro.optim import Adam, ClippedAdam
# Cuda GPU resources
torch.cuda.set_device(0)
torch.cuda.requires_grad = True
# fix random generator seed (for reproducibility of results)
np.random.seed(42)
# matplotlib style options
plt.style.use('ggplot')
%matplotlib inline
plt.rcParams['figure.figsize'] = (12, 8)
# Reading files from the local drive
dfv1 = pd.read_csv('//mnt//md0//data_vishnu//Pyro_Code//Revision_1_LDA//Version_1_Training_LDA.csv')
dfv2 = pd.read_csv('//mnt//md0//data_vishnu//Pyro_Code//Revision_1_LDA//Version_2_Training_LDA.csv')
dfv3 = pd.read_csv('//mnt//md0//data_vishnu//Pyro_Code//Revision_1_LDA//Version_3_Training_LDA.csv')
```

The mode is encoded as a integer from 0 to 2, corresponding to: -

0 - Regular Car

1 - Private Autonomous Vehicle

2 - Shared Autonomous Vehicle

**Frequency Distribution of Mode Choice**

```
def desc_stats(data, title):
    print("Dataset size: ", len(data))
```

```
        data['Choice'].hist();
        plt.title(title)
        plt.xlabel('Mode ID (0-regular car, 1 - Private Autonomous Vehicle, 2- Shared Autonomous Vehicle)')
        plt.ylabel('Frequency')
        plt.xticks([0,1,2]);
        return
desc_stats(dfv1, "Mode Choice (Version_1)");
dfv1.describe()
desc_stats(dfv2, "Mode Choice (Version_2)");
dfv2.describe()
desc_stats(dfv3, "Mode Choice (Version_3)");
dfv3.describe()
```

# Concatenate Dataset

```
df = pd.concat([dfv1, dfv2, dfv3], ignore_index=True)
desc_stats(df, "Mode Choice (Combined)");
df.describe()
```

**Data Preparation**

For the Likert scale questions, we created dummy variables and for the open-ended questions, we extracted variable using Latent Dirichlet allocation Method.

```
def data_processing(data):
    """
    inp:
        dataframe to be processed
    desc:
        create dummy variables for the discrete variables
        standardise continuous variables
    out:
        processed dataframe
    """
    # standardize input features
    X_mean1 = data.iloc[:, [41, 42]].mean(axis=0)
    X_std1 = data.iloc[:, [41, 42]].std(axis=0)
    data.iloc[:, [41, 42]] = (data.iloc[:, [41, 42]] - X_mean1) / X_std1
    X_mean2 = data.iloc[:, 78: 120].mean(axis=0)
    X_std2 = data.iloc[:, 78: 120].std(axis=0)
    data.iloc[:, 78: 120] = (data.iloc[:, 78: 120] - X_mean2) / X_std2
    # Converting the ordered Attitudinal Responses into Dummy Variables
    peu_names = ["AV_LeaEa", "AV_WrkEa", "AV_SklEa", "AV_UseEa"]
    pu_names  = ["AV_TrNed", "AV_OthAc", "AV_DeAcc", "AV_ReStr", "AV_UsImp"]
    psr_names = ["AV_WoSaf", "AV_MaFal"]
    ppr_names = ["AV_PeInf", "AV_UsInf", "AV_ShInf"]
    tr_names  = ["AV_Depe", "AV_Reli", "AV_Trus"]
    at_names  = ["AV_GIde", "AV_WIde", "AV_Plea"]
    X_PEU = np.concatenate([pd.get_dummies(data[x]) for x in peu_names], axis=1).astype("float32")
    X_PU  = np.concatenate([pd.get_dummies(data[x]) for x in pu_names], axis=1).astype("float32")
    X_PSR = np.concatenate([pd.get_dummies(data[x]) for x in psr_names], axis=1).astype("float32")
    X_PPR = np.concatenate([pd.get_dummies(data[x]) for x in ppr_names], axis=1).astype("float32")
    X_TR  = np.concatenate([pd.get_dummies(data[x]) for x in tr_names], axis=1).astype("float32")
    X_AT  = np.concatenate([pd.get_dummies(data[x]) for x in at_names], axis=1).astype("float32")
    # Grouping the Independent Variables into Different Sets
    mat = data.values
    X_SD = mat[:, [2,3,4,5,6,8,9,10,11,13,14,15,16,18,19,20,21,22,24,25,27,28,29]].astype("float32")
                                                                # Set of Socio-demographic Variables
    X_TC = mat[:, [30,31,33,34,35,36,37,38,39,46,47,48,49]].astype("float32")
                                                                # Set of Travel Characteristics
    X_FAV = mat[:, [56,57]].astype("float32")                   # Familiarity with Autonomous Vehicles
    X_SP = mat[:, 112:120].astype("float32")                    # SP Attribute Variables
```

```python
    X_TPEU = mat[:, 78:82].astype("float32")           # Topics for Perceived Ease of Use
    X_TPU  = mat[:, 82:89].astype("float32")           # Topics for Usefulness
    X_TPSR = mat[:, 89:95].astype("float32")           # Topics for Perceived Safety Risk
    X_TPPR = mat[:, 95:101].astype("float32")          # Topics for Perceived Privacy Risk
    X_TTR  = mat[:, 101:106].astype("float32")         # Topics for Trust
    X_TAT  = mat[:, 106:112].astype("float32")         # Topics for Attitudes
    bern = mat[:, -2].astype("int")                    # Question_type
    y = mat[:,-1].astype("int")
    # Concatenating the variables
    X_lk = np.concatenate([X_PEU, X_PU], axis=1)
    X_oe = np.concatenate([X_TPEU, X_TPU], axis=1)
    X_sd = np.concatenate([X_SD, X_TC, X_FAV, X_SP], axis=1)
    return y, X_lk, X_oe, X_sd, bern
y, X_lk, X_oe, X_sd, bern = data_processing(df)        # Processed data
bern1 = 1*(np.array([bern, bern]).transpose() == 1)
bern2 = 1*(np.array([bern, bern]).transpose() == 2)
bern3 = 1*(np.array([bern, bern]).transpose() == 3)
```

**Proposed Model**

**Pyro Model for the Combined Version**

```python
def model(X_ls, X_oe, X_sd, n_cat, bern1, bern2, bern3, K, obs=None):
    # Coefficients for the LS model (Version_1)
    alpha_ls1 = pyro.sample("alpha_ls1", dist.Normal(torch.zeros(K), torch.ones(K)))
    gamma_ls1 = pyro.sample("gamma_ls1", dist.Normal(torch.zeros(X_ls.shape[1], K),
                torch.ones(X_ls.shape[1], K)))
    # Coefficients for the LS model (Version_2)
    alpha_ls2 = pyro.sample("alpha_ls2", dist.Normal(torch.zeros(K), torch.ones(K)))
    gamma_ls2 = pyro.sample("gamma_ls2", dist.Normal(torch.zeros(X_ls.shape[1], K),
                torch.ones(X_ls.shape[1], K)))
    # Coefficients for the OE model
    alpha_oe = pyro.sample("alpha_oe", dist.Normal(torch.zeros(K), torch.ones(K)))
    gamma_oe = pyro.sample("gamma_oe", dist.Normal(torch.zeros(X_oe.shape[1], K),
                torch.ones(X_oe.shape[1], K)))
    with pyro.plate("data", X_ls.shape[0], use_cuda=True):
        y_att_ls1 = pyro.sample("y_att_ls1", dist.MultivariateNormal(alpha_ls1 + torch.matmul(X_ls,
                    gamma_ls1), torch.eye(K)), obs=None)
        y_att_ls2 = pyro.sample("y_att_ls2", dist.MultivariateNormal(alpha_ls2 + torch.matmul(X_ls,
                    gamma_ls2), torch.eye(K)), obs=None)
        y_att_oe = pyro.sample("y_att_oe", dist.MultivariateNormal(alpha_oe + torch.matmul(X_oe, gamma_oe),
                    torch.eye(K)), obs=None)
    y_att = bern1 * y_att_ls1 + bern2 * y_att_ls2 + bern3 * y_att_oe
    y_att = (y_att - torch.mean(y_att, dim=0))/torch.std(y_att, dim=0)
    X_Data = torch.zeros(X_sd.shape[0], X_sd.shape[1] + 2)
    X_Data[:, 0:-2] = X_sd
    X_Data[:, -2] = y_att[:, 0]
    X_Data[:, -1] = y_att[:, 1]
    # Coefficients for the Classification model
    alpha = torch.zeros(1, n_cat)
    alpha_1 = pyro.sample("alpha_1", dist.Normal(torch.zeros(n_cat-1), torch.ones(n_cat-1)))
    alpha[:, 1:] = alpha_1
    beta = torch.zeros(X_Data.shape[1], n_cat)
    beta_1 = pyro.sample("beta_1", dist.Normal(torch.zeros(X_Data.shape[1], n_cat-1),
             torch.ones(X_Data.shape[1], n_cat-1)))
    beta[:, 1:] = beta_1
    with pyro.plate("data_final", X_Data.shape[0], use_cuda=True):
        y = pyro.sample("y", dist.Categorical(logits= alpha + torch.matmul(X_Data, beta)), obs=obs)
    return y
```

**Preparing the tensors for the model (Proposed_Model)**

```python
n_cat = 3
K = 2
X_lk = torch.from_numpy(X_lk).float().cuda()
X_oe = torch.from_numpy(X_oe).float().cuda()
X_sd = torch.from_numpy(X_sd).float().cuda()
bern1 = torch.from_numpy(bern1).float().cuda()
bern2 = torch.from_numpy(bern2).float().cuda()
bern3 = torch.from_numpy(bern3).float().cuda()
y = torch.from_numpy(y).float().cuda()
```

**Inference using SVI**

```
%%time
```

**# Define guide function**

```python
guide = AutoDiagonalNormal(model)
```

**# Reset parameter values**

```python
pyro.clear_param_store()
```

**# Define the number of optimization steps**

```python
n_steps = 4000
```

**# Setup the optimizer**

```python
adam_params = {"lr": 0.01}
optimizer = ClippedAdam(adam_params)
```

**# Setup the inference algorithm**

```python
elbo = Trace_ELBO(num_particles=3)
svi = SVI(model, guide, optimizer, loss=elbo)
```

**# Do gradient steps**

```python
for step in range(n_steps):
    elbo = svi.step(X_lk, X_oe, X_sd, n_cat, bern1, bern2, bern3, K, y)
    if step % 500 == 0:
        print("[%d] ELBO: %.1f" % (step, elbo))
```

**Upon convergence, we can use the Predictive class to extract samples from posterior:**

```python
from pyro.infer import Predictive
def summary(samples):
    site_stats = {}
    for k, v in samples.items():
        site_stats[k] = {
            "mean": torch.mean(v, 0),
            "std": torch.std(v, 0),
            "5%": v.kthvalue(int(len(v) * 0.05), dim=0)[0],
            "95%": v.kthvalue(int(len(v) * 0.95), dim=0)[0],
        }
    return site_stats
predictive = Predictive(model, guide=guide, num_samples=2000,
                return_sites=("alpha_ls1", "gamma_ls1", "alpha_ls2", "gamma_ls2", "alpha_oe", "gamma_oe",
                "y_att_ls1", "y_att_ls2", "y_att_oe", "y_att", "alpha_1", "beta_1"))
samples = predictive(X_lk, X_oe, X_sd, n_cat, bern1, bern2, bern3, K, y)
pred_summary = summary(samples)
pred_summary.items()
predictions = pd.DataFrame({
    "alpha_ls1" : pred_summary["alpha_ls1"],
    "gamma_ls1" : pred_summary["gamma_ls1"],
    "alpha_ls2" : pred_summary["alpha_ls2"],
    "gamma_ls2" : pred_summary["gamma_ls2"],
    "alpha_oe"  : pred_summary["alpha_oe"],
    "gamma_oe"  : pred_summary["gamma_oe"],
    "y_att_ls1" : pred_summary["y_att_ls1"],
    "y_att_ls2" : pred_summary["y_att_ls2"],
    "y_att_oe"  : pred_summary["y_att_oe"],
    "alpha_1"   : pred_summary["alpha_1"],
```

```
    "beta_1"   : pred_summary["beta_1"]
})
predictions.head()
predictions.to_csv('coeff.csv')
```

We can now use the inferred posteriors to make predictions for the test set and compute the corresponding accuracy:

**Prediction Accuracy**

**Reading the files**

```
dft1 = pd.read_csv('//mnt//md0//data_vishnu//Pyro_Code//Revision_1_LDA//Version_1_Test_LDA.csv')
dft2 = pd.read_csv('//mnt//md0//data_vishnu//Pyro_Code//Revision_1_LDA//Version_2_Test_LDA.csv')
dft3 = pd.read_csv('//mnt//md0//data_vishnu//Pyro_Code//Revision_1_LDA//Version_3_Test_LDA.csv')
y_v1, X_lk_v1, X_oe_v1, X_sd_v1, bern_v1 = data_processing(dft1)
y_v2, X_lk_v2, X_oe_v2, X_sd_v2, bern_v2 = data_processing(dft2)
y_v3, X_lk_v3, X_oe_v3, X_sd_v3, bern_v3 = data_processing(dft3)
# Coefficients for the LS model- Ver_LK
alpha_ls1 = samples["alpha_ls1"].cpu()
gamma_ls1 = samples["gamma_ls1"].cpu()
alpha_ls1_hat=np.array([np.mean(b, axis=0) for b in alpha_ls1.detach().numpy().T])
gamma_ls1_hat=np.array([np.mean(b, axis=1) for b in gamma_ls1.detach().numpy().T]).T
# Coefficients for the LS model- Ver_LKOE
alpha_ls2 = samples["alpha_ls2"].cpu()
gamma_ls2 = samples["gamma_ls2"].cpu()
alpha_ls2_hat=np.array([np.mean(b, axis=0) for b in alpha_ls2.detach().numpy().T])
gamma_ls2_hat=np.array([np.mean(b, axis=1) for b in gamma_ls2.detach().numpy().T]).T
# Coefficients for the OE model- Ver_OE
alpha_oe = samples["alpha_oe"].cpu()
gamma_oe = samples["gamma_oe"].cpu()
alpha_oe_hat=np.array([np.mean(b, axis=0) for b in alpha_oe.detach().numpy().T])
gamma_oe_hat=np.array([np.mean(b, axis=1) for b in gamma_oe.detach().numpy().T]).T
# Coefficients for the choice
alpha_1 = samples["alpha_1"].cpu()
beta_1 = samples["beta_1"].cpu()
alpha_1_hat=np.array([np.mean(b) for b in alpha_1.detach().numpy().T])
beta_1_hat=np.array([np.mean(b, axis=1) for b in beta_1.detach().numpy().T])
y_att_ls1 = samples["y_att_ls1"].cpu()
y_att_ls2 = samples["y_att_ls2"].cpu()
y_att_oe = samples["y_att_oe"].cpu()
y_att_ls1_hat=np.array([np.mean(b, axis=1) for b in y_att_ls1.detach().numpy().T]).T
y_att_ls2_hat=np.array([np.mean(b, axis=1) for b in y_att_ls2.detach().numpy().T]).T
y_att_oe_hat=np.array([np.mean(b, axis=1) for b in y_att_oe.detach().numpy().T]).T
bern1 = bern1.cpu()
bern2 = bern2.cpu()
bern3 = bern3.cpu()
y_att_hat =  bern1.detach().numpy() *  y_att_ls1_hat + bern2.detach().numpy() * y_att_ls2_hat +
             bern3.detach().numpy() * y_att_oe_hat
np.savetxt("y_att_pred.csv", y_att_hat, delimiter=',')
```

**Accuracy for Version_1 (Test Set)**

```
y_hat_v1 = [None]*len(y_v1)
X_Data = np.zeros((X_sd_v1.shape[0], X_sd_v1.shape[1] + 2))
X_Data[:, 0:-2] = X_sd_v1
alpha_hat = np.zeros(n_cat)
alpha_hat[1:] = alpha_1_hat
beta_hat = np.zeros((n_cat, X_Data.shape[1]))
beta_hat[1:, :] = beta_1_hat
y_att_v1 = [[None]*2]*len(y_v1)
for i in range(len(y_v1)):
```

```python
        y_att_v1[i] = alpha_ls1_hat + np.dot(X_lk_v1[i], gamma_ls1_hat)
        X_Data[i, -2] = y_att_v1[i][0]
        X_Data[i, -1] = y_att_v1[i][1]
        y_hat_v1[i] = alpha_hat + np.dot(beta_hat, X_Data[i])
# opening the csv file in 'w+' mode
file = open('utilities_comb_lk.csv', 'w+', newline ='')
# writing the data into the file
with file:
    write = csv.writer(file)
    write.writerows(y_hat_v1)
y_hat_v1 = np.argmax(y_hat_v1, axis=1)
print("predictions:", y_hat_v1)
print("true values:", y_v1)
print(np.unique(y_hat_v1))
print(np.unique(y_v1))
# evaluate prediction accuracy
print("Accuracy:", 1.0*np.sum(y_hat_v1 == y_v1) / len(y_v1))
```

**Accuracy for Version_2 (Test Set)**

```python
y_hat_v2 = [None]*len(y_v2)
X_Data = np.zeros((X_sd_v2.shape[0], X_sd_v2.shape[1] + 2))
X_Data[:, 0:-2] = X_sd_v2
y_att_v2 = [[None]*2]*len(y_v2)
for i in range(len(y_v2)):
    y_att_v2[i] = alpha_ls2_hat + np.dot(X_lk_v2[i], gamma_ls2_hat)
    X_Data[i, -2] = y_att_v2[i][0]
    X_Data[i, -1] = y_att_v2[i][1]
    y_hat_v2[i] = alpha_hat + np.dot(beta_hat, X_Data[i])
# opening the csv file in 'w+' mode
file = open('utilities_comb_lkoe.csv', 'w+', newline ='')
# writing the data into the file
with file:
    write = csv.writer(file)
    write.writerows(y_hat_v2)
y_hat_v2 = np.argmax(y_hat_v2, axis=1)
print("predictions:", y_hat_v2)
print("true values:", y_v2)
print(np.unique(y_hat_v2))
print(np.unique(y_v2))
# evaluate prediction accuracy
print("Accuracy:", 1.0*np.sum(y_hat_v2 == y_v2) / len(y_v2))
```

**Accuracy for Version_3 (Test Set)**

```python
y_hat_v3 = [None]*len(y_v3)
X_Data = np.zeros((X_sd_v3.shape[0], X_sd_v3.shape[1] + 2))
X_Data[:, 0:-2] = X_sd_v3
y_att_v3 = [[None]*2]*len(y_v3)
for i in range(len(y_v3)):
    y_att_v3[i] = alpha_oe_hat + np.dot(X_oe_v3[i], gamma_oe_hat)
    X_Data[i, -2] = y_att_v3[i][0]
    X_Data[i, -1] = y_att_v3[i][1]
    y_hat_v3[i] = alpha_hat + np.dot(beta_hat, X_Data[i])
# opening the csv file in 'w+' mode
file = open('utilities_comb_oe.csv', 'w+', newline ='')
# writing the data into the file
with file:
    write = csv.writer(file)
    write.writerows(y_hat_v3)
```

```
y_hat_v3 = np.argmax(y_hat_v3, axis=1)
print("predictions:", y_hat_v3)
print("true values:", y_v3)
print(np.unique(y_hat_v3))
print(np.unique(y_v3))
# evaluate prediction accuracy
print("Accuracy:", 1.0*np.sum(y_hat_v3 == y_v3) / len(y_v3))
```

**Accuracy for Version_1 (Training Set)**

```
y, X_lk, X_oe, X_sd, bern = data_processing(dfv1)
y_hat = [None]*len(y)
X_tr_Data = np.zeros((X_sd.shape[0], X_sd.shape[1] + 2))
X_tr_Data[:, 0:-2] = X_sd
alpha_tr_hat = np.zeros(n_cat)
alpha_tr_hat[1:] = alpha_1_hat
beta_tr_hat = np.zeros((n_cat, X_tr_Data.shape[1]))
beta_tr_hat[1:, :] = beta_1_hat
y_att_tr = [[None]*2]*len(y)
for i in range(len(y)):
    y_att_tr[i] = alpha_ls1_hat + np.dot(X_lk[i], gamma_ls1_hat)
    X_tr_Data[i, -2] = y_att_tr[i][0]
    X_tr_Data[i, -1] = y_att_tr[i][1]
    y_hat[i] = alpha_tr_hat + np.dot(beta_tr_hat, X_tr_Data[i])
# opening the csv file in 'w+' mode
file = open('utilities_comb_lk_tr.csv', 'w+', newline ='')
# writing the data into the file
with file:
    write = csv.writer(file)
    write.writerows(y_hat)
y_hat = np.argmax(y_hat, axis=1)
print("predictions:", y_hat)
print("true values:", y)
print(np.unique(y_hat))
print(np.unique(y))
# evaluate prediction accuracy
print("Accuracy:", 1.0*np.sum(y_hat == y) / len(y))
```

**Accuracy for Version_2 (Training Set)**

```
y, X_lk, X_oe, X_sd, bern = data_processing(dfv2)
y_hat = [None]*len(y)
X_tr_Data = np.zeros((X_sd.shape[0], X_sd.shape[1] + 2))
X_tr_Data[:, 0:-2] = X_sd
alpha_tr_hat = np.zeros(n_cat)
alpha_tr_hat[1:] = alpha_1_hat
beta_tr_hat = np.zeros((n_cat, X_tr_Data.shape[1]))
beta_tr_hat[1:, :] = beta_1_hat
y_att_tr = [[None]*2]*len(y)
for i in range(len(y)):
    y_att_tr[i] = alpha_ls2_hat + np.dot(X_lk[i], gamma_ls2_hat)
    X_tr_Data[i, -2] = y_att_tr[i][0]
    X_tr_Data[i, -1] = y_att_tr[i][1]
    y_hat[i] = alpha_tr_hat + np.dot(beta_tr_hat, X_tr_Data[i])
# opening the csv file in 'w+' mode
file = open('utilities_comb_lkoe_tr.csv', 'w+', newline ='')
# writing the data into the file
with file:
    write = csv.writer(file)
    write.writerows(y_hat)
y_hat = np.argmax(y_hat, axis=1)
```

```python
print("predictions:", y_hat)
print("true values:", y)
print(np.unique(y_hat))
print(np.unique(y))
# evaluate prediction accuracy
print("Accuracy:", 1.0*np.sum(y_hat == y) / len(y))
```

**Accuracy for Version_3 (Training Set)**

```python
y, X_lk, X_oe, X_sd, bern = data_processing(dfv3)
y_hat = [None]*len(y)
X_tr_Data = np.zeros((X_sd.shape[0], X_sd.shape[1] + 2))
X_tr_Data[:, 0:-2] = X_sd
alpha_tr_hat = np.zeros(n_cat)
alpha_tr_hat[1:] = alpha_1_hat
beta_tr_hat = np.zeros((n_cat, X_tr_Data.shape[1]))
beta_tr_hat[1:, :] = beta_1_hat
y_att_tr = [[None]*2]*len(y)
for i in range(len(y)):
    y_att_tr[i] = alpha_oe_hat + np.dot(X_oe[i], gamma_oe_hat)
    X_tr_Data[i, -2] = y_att_tr[i][0]
    X_tr_Data[i, -1] = y_att_tr[i][1]
    y_hat[i] = alpha_tr_hat + np.dot(beta_tr_hat, X_tr_Data[i])
# opening the csv file in 'w+' mode
file = open('utilities_comb_oe_tr.csv', 'w+', newline ='')
# writing the data into the file
with file:
    write = csv.writer(file)
    write.writerows(y_hat)
y_hat = np.argmax(y_hat, axis=1)
print("predictions:", y_hat)
print("true values:", y)
print(np.unique(y_hat))
print(np.unique(y))
# evaluate prediction accuracy
print("Accuracy:", 1.0*np.sum(y_hat == y) / len(y))
```

# Appendix E (Python Code for the Mapping of Responses)

```python
import numpy as np
import pandas as pd
from sklearn.preprocessing import OneHotEncoder
np.random.seed(42)
```

**Function for Data Pre-processing**

```python
def data_preprocessing(len_dataset, ques_size):
    """
    def:
        pre-processing of data
    inp:
        len_dataset- length of the dataset
        ques_size- array with number of topics/levels per question
    out:
        dataset with onehotencoded variables
    """
    ### Creating the dummy Dataset
    len_dataset = len_dataset
    num_ques = len(ques_size)
    X_Act = np.array(np.zeros((len_dataset, num_ques)))
    for i in range(num_ques):
        X_Act[:, i] = np.random.choice(a=ques_size[i], size=(len_dataset, 1), p=[1/ques_size[i]]*ques_size[i]).T
    var_name = ['X_' + str(i + 1) for i in range(num_ques)]
    X_df = pd.DataFrame(data=X_Act, columns=var_name)
    ### OneHotEncoding of the Likert scale responses
    enc = OneHotEncoder(handle_unknown='ignore')
    enc_name = [0]*num_ques
    for k in range(num_ques):
        enc_name[k] = pd.DataFrame(enc.fit_transform(X_df[[var_name[k]]]).toarray())
        enc_name[k].columns = ['X_'+str(k+1)+'_' + str(i+1) for i in range(ques_size[k])]
        if k > 0:
            enc_name[k] = enc_name[k-1].join(enc_name[k])
    return enc_name[num_ques-1]
```

**Function for Gibbs Sampling**

```python
def discrete_gibbs(Data, gamma_df, gamma_inv, alpha_df, ques_size, n_iter, n_warm):
    """
    def:
        function that performs Gibbs Sampling for discrete variables
    inp:
        Data- Dataset
        gamma_df- Dataset of coefficients
        gamma_inv- Dataset of the inverse coefficients
        alpha_df- Dataset of constants
        ques_size- array with number of topics/levels per question
        n_iter- number of iterations
        n_warm- number of warmup iterations to be discarded
    out:
        Dataset with sampled values
    """
    num_ques = np.sum(ques_size)
    y_arr = ['y_att_1', 'y_att_2']
    np_arr = np.array([[[0.000]*(n_iter+n_warm)]*num_ques]*len(Data))
    for j in range(n_iter):
        print("Iterations: ", j)
        for index, row in Data.iterrows():
            n_dims = 0
```

```python
            for k in range(len(ques_size)):
                V_na = ['X_'+str(k+1)+'_' + str(i+1) for i in range(ques_size[k])]
                V_nn = V_na + y_arr
                fir_pa = np.array([row['y_att_1'], row['y_att_2']]) - alpha_df - np.matmul(np.array(gamma_df.drop(columns=V_na)), np.array(row.drop(V_nn, axis=0)))
                inv_mc = np.array(gamma_inv.loc[V_na, :].T)
                x_value = np.matmul(fir_pa, inv_mc)
                sum_val = np.sum(np.abs(x_value))
                x_val = np.abs(x_value)/sum_val
                x_val = np.random.dirichlet(x_val)
                for m in range(ques_size[k]):
                    np_arr[index, n_dims + m, j] = x_val[m]
                n_dims += ques_size[k]
        if j == n_iter/4:
            print("25% iteration complete")
        elif j == n_iter/2:
            print("50% iteration complete")
        elif j == 3*n_iter/4:
            print("75% iteration complete")
        elif j == n_iter-1:
            print("Iteration complete")
    return np_arr
```

**Execution**

```python
def dis_sample(Dataset, gamma_df, gamma_inv, alpha_df, ques_size, num_iter, num_warm):
    """
    def:
        Perform discrete Gibbs sampling on a dataset
    inp:
        Dataset containing y_variables
        gamma_df- data frame with the coefficients
        gamma_inv- data frame with the inverse of the coefficients
        ques_size- array with the number of topics/levels per question
        num_iter- total number of iterations
        num_warm- number of warm-ups
    out:
        Sampled dataset
    """
    len_dataset = len(Dataset)
    Exp_Var = pd.DataFrame(data_preprocessing(len_dataset,ques_size))
    Dataset = Dataset.join(Exp_Var)
    # Gibbs Sampling
    Dataset = discrete_gibbs(Dataset, gamma_df, gamma_inv, alpha_df, ques_size, num_iter, num_warm)
    Data_1 = Dataset[:, :, num_warm:]
    Data_2 = np.mean(Data_1, axis=2)
    print(Data_2.shape)
    X_Variables = np.mean(Data_2, axis=0)
    print(X_Variables)
    return Dataset

# For Likert scale questions, set values
num_questions = 9
ques_size = [5]*num_questions
Dataset = pd.read_csv("y_att_pred_v1.csv", header=None)
Dataset.columns = ['y_att_1', 'y_att_2']
print(len(Dataset))
gamma_v  = pd.read_csv("Gamma_ls2.csv", header=None)
alpha_v = np.array([0.6820, -0.7710])         # Update values
col_names = []
for j in range(num_questions):
    col_names.extend(['X_'+str(j+1)+'_' + str(i+1) for i in range(ques_size[j])])
```

```python
gamma_v.columns = col_names
for i in range(num_questions):
    V_na = ['X_'+str(i+1)+'_' + str(j+1) for j in range(ques_size[i])]
    gamma_v1 = gamma_v[V_na]
    if i == 0:
        gamma_inv = pd.DataFrame((gamma_v1.T).dot(np.linalg.inv(gamma_v1.dot(gamma_v1.T))))
        gamma_inv.index = V_na
    else:
        gamma_inv1 = pd.DataFrame((gamma_v1.T).dot(np.linalg.inv(gamma_v1.dot(gamma_v1.T))))
        gamma_inv1.index = V_na
        gamma_inv = gamma_inv.append(gamma_inv1)
gamma_inv.columns = ['y_att_1', 'y_att_2']
Data = dis_sample(Dataset, gamma_v, gamma_inv, alpha_v, ques_size, 2000, 400)
Data_1 = Data[:, :, 50:]
print(Data_1.shape)
Data_2 = np.mean(Data_1, axis=2)
print(Data_2.shape)
X_Variables = np.mean(Data_2, axis=0)
print(X_Variables)
import csv
with open('Data_ls2_2000.csv', 'w', newline='') as csvfile:
    writer = csv.writer(csvfile, delimiter=',')
    writer.writerows(Data_2)
```